\documentclass[12pt]{article}
\usepackage{graphicx}
\usepackage{latexsym,amsfonts, amssymb,epsfig,amsmath,cite}
\usepackage[english]{babel}

\newtheorem{thm}{Theorem}

\newtheorem{st}[thm]{Statement}
\newtheorem{rmk}{Remark}

\begin{document}

\begin{center}{\large\bf Stability and dynamical features of solitary wave
so\-lu\-ti\-ons for a hy\-d\-ro\-dy\-namic-type system taking into
account non-local effects}

\vspace{8mm}

%\tnotetext[t1]{Stability of the solitary wave solutions to the
%Nonl. H-dyn System}

{\it Vsevolod Vladimirov$^1$, Czes\l aw M\c{a}czka$^1$,
Artur Sergyeyev$^2$, Sergiy Skurativskyi$^3$\\
$^1$Faculty of Applied Mathematics,
AGH University of Science and Technology,\\
Mickiewicz Avenue 30, 30-059 Krak\'{o}w, Poland\\
$^2$Mathematical Institute,
Silesian University in Opava, \\
Na Rybn\'\i{}\v cku 1, 74601 Opava, Czech Republic\\
$^3$Subbotin Institute for Geophysics of NAS of Ukraine,\\
32 Palladina av., 03142 Kyiv, Ukraine\\[3mm]
E-mail: {\tt vsevolod.vladimirov@gmail.com, czmaczka@agh.edu.pl},\\
{\tt artur.sergyeyev@gmail.com, skurserg@gmail.com}}
\end{center}

\vspace{4mm}

\begin{abstract}
We consider a hydrodynamic-type system of balance equations
for mass and momentum closed by the dynamical equation of state
taking into account the effects of spatial nonlocality.
We study higher symmetries and local conservation laws for this system
and establish its nonintegrability for the generic values of parameters.
A system of ODEs obtained from the system under study through the group
theory reduction is investigated. The reduced system is shown to
possess a family of the homoclinic solutions describing solitary
waves of compression and rarefaction. The waves of compression are
shown to be unstable. On the contrary, the waves of rarefaction are
likely to be stable. Numerical simulations reveal some peculiarities
of solitary waves of rarefaction, and, in particular, the recovery of
their shape after the collisions
\end{abstract}

\vspace{4mm}


\section{Introduction}

This paper deals with soliton-like traveling wave (TW) solutions to
some nonlinear evolutionary PDEs. The TW solutions play an important
role in mathematical physics. They appear in the models of  various
transport phenomena, including the shallow water equation
\cite{KdV}, the lithosphere model \cite{Lund}, the nerve axon model
\cite{FitzHugh-Nagumo,Evans,Feroe},  models of
combustion theory \cite{Zel'dovich}, mathematical biology
\cite{KPP},  and active dissipative media models
\cite{Nakoryakov,Demekhin-Shkadov} (for further examples see \cite{Kersner,Sandstede} and references therein). Among the great variety of nonlinear wave
solutions, perhaps the most important are {\it solitons}, supported, in particular,
by the celebrated Korteweg--de Vries (KdV) equation.
After the  the discovery of integrability of
the KdV equation in the second half of the XX-th century, the study of { solitons}
attracted attention of a great many authors, see e.g.\ \cite{KdV} and references therein. The
unusual features of soliton solutions, in particular their particle-like
collisions and the existence of a broad set of smooth Cauchy data
asymptotically turning into a finite number of solitons in the course of time evolution,
have usually been attributed to the complete integrability
of the KdV equation, and certain attendant properties, such as the infinite
symmetry and existence of an infinite number of local conservation
laws \cite{Olver,BlumanCheviakov}. Later on it was observed that solutions to the equations which are not completely integrable manifest features similar to those demonstrated by ''true'' solitons.
A prominent example of this is provided by the so-called
{\it compactons}, supported by the equations of the $K(m,n)$ family, introduced
by J. Hyman and Ph. Rosenau in 1993 \cite{HymanRosenau}. The non-integrability of a generic member of the class of
Hyman--Rosenau $K(m,n)$ equations stems {\em inter alia} from the fact that only a finite number of higher symmetries
is known and only four local conservation laws have been obtained to date \cite{OverRosenau}
(it was rigorously proved recently that the $K(m,n)$ equations
for $m=n$ and $m\neq  -2, -1/2, 0, 1$ have just four conservation laws and no higher symmetries \cite{Vodova}).
The family of non-integrable evolutionary equations possessing stable localized solutions
with soliton features is not exhausted by the generic $K(m,n)$ equations. Other
non-integrable models with similar features of soliton-like
solutions are presented in \cite{Makhan'kov,Makhan'kov2,Nesterenko2,Nesterenko,PikovskyRos,RosPikovsky,Kurkiny}.

The main goal of the present paper is to demonstrate the existence of
stable soliton-like TW solutions in a hydrodynamic-type model taking
into account the effects of spatial nonlocality.  The study of higher symmetries
and local conservation laws admitted by the system under study reveals
its non-integrability for generic values of parameters.
Nevertheless, the solitary
waves supported by this system restore their shapes after the mutual collisions.
Furthermore, a proper choice of the Cauchy data leads to the creation of a chain
of solitary waves moving with distinct velocities.

The paper is organized as follows. In section 2 we introduce our
basic system of PDEs and study its higher symmetries and local
conservation laws. Using these results,
we establish non-integrability of this system
for generic values of parameters. In section 3 we give a geometric insight into
the structure of the phase space of the dynamic system obtained
through the symmetry-based reduction of the initial system, and
formulate the conditions which guarantee that the reduced system
possesses a one-parameter family of homoclinic solutions
corresponding to the solitary wave regimes. In section 4 we study the
perturbed solitary wave solutions and estimate the essential spectrum
of the linearized problem. In section 5 we perform a
numerical study of stability for the solitary waves of rarefaction
(dark solitons) using the Evans function and investigate some peculiarities
of the solitary waves, such as the dependence of their maximal depth and the
effective width on the wave velocity, evolution of the initial
non-solitonic perturbation and the behavior of waves in the course of
their collisions. Finally, in section 6 we discuss the results
obtained and give an outline of further research.

%\end{document}

\section{The model system and its symmetry properties}

It is well known that the features of TW depend in  essence on
dispersive and nonlinear properties of physical media
\cite{Witham,Bhatnagar,KdV,vsandan1,makar1,peerlart,Danbook}. Let us stress that in many cases
nonlocal effects caused by internal structure of the media can
also play significant role in formation and
evolution of wave patterns. The presence of internal structure causes various effects
such as the fragmentation of smooth initial perturbations and
intensification of shock fronts in the models of heterogeneous media
\cite{Nakoryakov,Nigmatulin}, soliton and compacton features of the
block media models \cite{KdV,Nesterenko,HymanRosenau}, and many
others.

  Unfortunately, there exists no universal model describing structured
media in a sufficiently wide range of the values of parameters. Below
we introduce a model system taking into account the nonlocal
effects related to the presence of internal structure (cf.\
\cite{Peerlings,VLADKI2004,VKUZOV12}). This model applies when
the ratio of the characteristic size $d$ of elements
of the structure to the characteristic wavelength $\lambda$ of the
wave pack is much smaller than unity and therefore the continual
approach is still valid but the said ratio is not small enough to ignore
the presence of internal structure.
%2
As it is shown in \cite{vakhkul,vakhdan}, the balance equations for
mass and momentum of media with internal structure in the long-wave
approximation still retain their classical form, which in the
one-dimensional case reads
%(cf. with
%\cite{vsan99}):
\begin{equation}\label{Eq:balanceq}
\left\{
\begin{array}{ll}
 u_t+p_x=0, \\
 \rho_t+\rho ^2 u_x =0.
\end{array} \right.
\end{equation}
Here $u$ denotes the mass velocity, $p$ is the pressure, $\rho$ is the
density, t is the time, $x$ is the mass (Lagrangian) coordinate related to
the conventional spatial coordinate $x_e$  as follows:
 \[ x=\int^{x_e}{\rho(t,\,\xi)\,d\,\xi}.
\]
The subscripts $t$ and $x$ denote partial derivatives with respect to indicated
variables.

Thus, the entire information about the presence of
structure in this approximation is contained in a dynamical equation
of state (DES) which should be incorporated into the system
(\ref{Eq:balanceq}) in order to make it closed.
In general, DES for structured media
manifesting the nonlocal features takes the form of an integral
equation \cite{rudyak,zubti} relating the generalized thermodynamical
flow $J$ and the generalized thermodynamical force $X$ causing this
flow:
\begin{equation}\label{Eq:intgen}
J=\int_{-\infty}^{t}\left[\int_{R}K(t,t';\,x,x')X(t',x')\,dx'\right]\,dt'.
\end{equation}
Here $K(t,t';\,x,x')$ is the kernel of nonlocality. The function $K$ can
be calculated by solving the dynamical problem of
structure elements interaction; however, such calculations are
extremely difficult. Therefore, in practice one uses, as a rule,
some model kernel describing well enough the main properties of
the nonlocal effects and, in particular, the fact that these
effects vanish rapidly as $|t-t'|$ and $|x-x'|$ grow. This property
could be used for passing from the integral equation
(\ref{Eq:intgen}) to a purely differential equation.

% ***************DES*********Spatial nonlocality********

One of the simplest equations of state taking into account the effects of
spatial nonlocality takes the form
\begin{equation}\label{Eq:intspat}
p= \hat{\sigma}\, \int_{-\infty}^{+\infty}K_1\left(x,\,x'
\right)\,\rho^{n}(t,\,x')\,dx',
\end{equation}
where $K_1\left(x,\,x' \right)=\exp\left[-{(x-x')^2}/{l^2}\right]$
is the kernel accounting for a purely spatial nonlocality. This model is attributed in
\cite{Peerlings} to the situation when the density of a medium  changes abruptly from point to point,
as this is the case  with an elastic body containing the microcracks or low density solid inclusions.

Using the fact that the function
$\exp{\left[-(x-x')^2/{l^2}\right]}$ extremely quickly approaches
 zero as $|x-x'|$ grows, one can replace the function  $\rho^{n}(t,\,x')$
 by several terms of its power series expansion,
\[
\rho^{n}(t,\,x')=\rho^{n}(t,\,x)+[\rho^{n}(t,\,x)]_{x}\frac{x'-x}{1!}+
[\rho^{n}(t,\,x)]_{xx}\frac{\left(x'-x\right)^2}{2!}+O(|x-x'|^{3}),
\]
obtaining in this way the following approximate flow-force relation:
\begin{equation}\label{Eq:tspat}
p= c_{0}\,\rho^{n}(t,\,x)+c_{2}\,\left[\rho^{n}(t
,x)\right]_{xx}.
\end{equation}
Here
\[
c_{0}=\hat{\sigma}\, l\int_{-\infty}^{+\infty}e^{-\tau^{2}}d\tau=\hat{\sigma}\,l\sqrt{\pi},
\qquad
c_{2}=\hat{\sigma}\,\frac{l^{3}}{2}\int_{-\infty}^{+\infty}\tau^{2}e^{-\tau^{2}}d\tau=\hat{\sigma}\, \frac{l^3\,\sqrt{\pi}}{4}.
\]
so both of the coefficients are positive. Choosing the kernel in the
form
\[
K(x',\,x)=\hat{\sigma}\,\left[\mu+\alpha\, (x-x')^2
\right]\,\exp\left[-{(x-x')^2}/{l^2}\right]
\]
and following the procedure outlined above, we obtain the dynamical equation
of state
\[
p= \hat{\sigma}\,\sqrt{\pi}\,\left\{\rho^{n}(t,\,x)
\,l\,\left(\mu+\frac{\alpha}{2}\,l^2 \right)+\,\left[\rho^{n}(t
,x)\right]_{xx}l^3\,\left(\frac{\mu}{4}+\frac{3}{8}\alpha \,l^2
\right)\right\}=
\]
\[
=\tilde{c_{0}}\,\rho^{n}(t',x)+\tilde{c_{2}}\,\left[\rho^{n}(t
,x)\right]_{xx}.
\]
For
$ \alpha=-1, \qquad {l^2}/{2}<\mu<{3\,l^2}/{2}$
$\tilde{c_{0}}$ remains positive while $\tilde{c_{2}}$ becomes negative. Note that
a well-established linear strain-stress dependence corresponding to the above situation
is presented in \cite{AriEringen}. Nonlinear strain-stress relationship of this type can
be used for the description of wave movements in the pre-stressed structured medium \cite{Nesterenko}.

Thus, we consider the model system
\begin{equation}\label{Vladimirov:mainspat}
\left\{
\begin{array}{l}
u_{t}+\beta\rho^{\nu+1}\rho_{x}+\gamma\left[\rho^{\nu+1}
\rho_{xxx}+3(1+\nu)\rho^{\nu}\rho_{x}\rho_{xx}+
\nu(1+\nu)\rho^{\nu-1}\rho^{3}_{x}\right]=0, \\\\
\rho_{t}+\rho^{2}u_{x}=0,\end{array} \right.
\end{equation}
assuming that $\beta>0$ and $\gamma\neq 0$.
%*****************************************************************
%**********************************ATTENTION!!!!!*******************************
% ************************A FEW  INTRODUCTORY WORDS***********************
%****************************ARE NEEDED IN THIS PLACE****************
%*****************************************************************
%*****************************************************************
%*****************************************************************

Let us turn to the study of integrability properties of (\ref{Vladimirov:mainspat}).
We shall use the integrability test based on existence of higher symmetries, see e.g.
\cite{Fokas, mikshab} for details,
because it allows to detect both integrability via the inverse scatttering transform and
linearizability through a suitably chosen transformation and, unlike, say, the search for a
Lax representation, verifying the existence of higher symmetries is an algorithmic procedure.

First of all, note that for $\nu=-3$ the change of variables
$u=u$, $r=1/\rho$ linearizes this system into
\begin{equation}\label{Vladimirov:mainspat1}
%\left\{
\begin{array}{ll}
u_{t}+\beta r_{x}+\gamma r_{xxx}=0,\\
r_{t}-u_{x}=0,\end{array}
\end{equation}
so for $\nu=-3$ the system (\ref{Vladimirov:mainspat}) is integrable (more precisely,
in this case it is $C$-integrable, i.e.,
linearizable).

Now turn to the case $\nu\neq -3$ and perform the following change of variables:
$v=\rho$, $w=u+c \rho^{(\nu-1)/2}\rho_x$, where $c=\sqrt{\gamma}$.
It converts our system into the following
\begin{equation}\label{Vladimirov:mainspat2}
%\left\{
\begin{array}{ll}
v_{t}=-c v^{(\nu+3)/2} v_{xx}-(c/2) v^{(\nu-1)/2} (\nu-1) v_x^2-v^2 w_x,\\
w_{t}=c v^{(\nu+3)/2} u_{xx}-(\nu+3) (c^2/2) (v^{\nu} v_x^2)_x
+c (\nu+3) v^{(\nu+1)/2} u_x v_x/2-b v^{\nu+1} v_x.\end{array}
\end{equation}

The matrix at the second derivatives in (\ref{Vladimirov:mainspat2}) has the form
\begin{equation}\label{symbol}
A=\left(
\begin{array}{rl}
- c v^{(\nu+3)/2} & 0\\
-c^2 (\nu+3) v^\nu v_x & c v^{(\nu+3)/2}\end{array}\right)
\end{equation}
with two distinct eigenvalues $\pm \lambda$, where $\lambda=c v^{(\nu+3)/2}$.

Then a necessary condition for (\ref{Vladimirov:mainspat2}) to admit
local higher, also known as generalized \cite{Olver},
symmetries (see e.g. \cite{Ibragimov,Olver, BlumanCheviakov} and references therein for details)
of order greater than two is (see e.g.\ \cite{mikshab})
that the quantity $\lambda^{-1/2}$ is a (local) conserved
density for (\ref{Vladimirov:mainspat2}), i.e., there exists a $\sigma$
depending on $x,t,v,w$ and a finite number of $x$-derivatives of $v$ and $w$ such that
$D_t(\lambda^{-1/2})=D_x(\sigma)$ by virtue of (\ref{Vladimirov:mainspat2}).
However, it is readily checked that $\lambda^{-1/2}$ is a conserved
density for (\ref{Vladimirov:mainspat2}) only if $\nu=-3$ or $\nu=1$. Therefore, for
$\nu\neq -3, 1$ the system (\ref{Vladimirov:mainspat2})
has local higher symmetries of order at most two. These are easily seen to have been exhausted by
the Lie point symmetries (more precisely, by symmetries equivalent to the Lie point ones).

Going back to (\ref{Vladimirov:mainspat}) and making use
of the transformation properties of higher symmetries, see e.g.
\cite{Ibragimov,BlumanCheviakov},
we conclude that (\ref{Vladimirov:mainspat})
has no symmetries of order greater than three. The computation of symmetries of order not greater
than three shows that for $\nu\neq -3, 1$ the only local higher
symmetries of (\ref{Vladimirov:mainspat2}) are (equivalent to)
the Lie point ones:
\begin{equation}\label{symm}
\begin{split}
Q_1 &=\partial/\partial u,\\
Q_2 &=\partial/\partial x,\\
Q_3 &=\partial/\partial t,\\
Q_4 &=(\nu+3)t\partial/\partial t-(\nu+1)u\partial/\partial u-2\rho \partial/\partial \rho.\\
\end{split}
\end{equation}
Thus, for $\nu\neq -3, 1$ the system (\ref{Vladimirov:mainspat}) is {\em non-integrable}, at least
in the sense of absence of genuinely higher symmetries, which is a strong indicator of non-integrability in any
other sense as well, cf. \cite{Fokas, mikshab, Olver}.
%Nevertheless, the system (\ref{Vladimirov:mainspat})
%possesses solitary wave solutions which inherit some properties of
%the ``true" solitons, as will be shown in the subsequent sections.\looseness=-1

The Lie point symmetries (\ref{symm}) have a clear physical interpretation:
the first three are translations w.r.t. independent variables and the dependent
variable $u$ while the last one is the scaling symmetry. Somewhat
surprisingly, this scaling symmetry does not involve $x$.

For the generic values of parameters these are the only (be
it Lie point or higher) symmetries admitted by the system
(\ref{Vladimirov:mainspat}). However, for special values of
parameters additional symmetries may emerge.

For instance, if we concentrate on the physically relevant case of
$\nu\,\geq\,0$, we easily find that for $\nu=1$ there appears an
additional nonlocal symmetry. Namely, consider the extended (also
known as potential, see e.g. \cite{BlumanCheviakov} for details)
system which consists of (\ref{Vladimirov:mainspat}) and the
equations for the potential $q$ of the conservation law given by the
second equation of (\ref{Vladimirov:mainspat}), that is,
\begin{equation}\label{potl}
q_x=1/\rho,\quad q_t=u.
\end{equation}
The said extended system for $\nu=1$ possesses the following
additional symmetry:
\[
Q_5=-t^2 \partial/\partial t+(t u-q)\partial/\partial u+  t \rho
\partial/\partial \rho+(t^2 u- t q)\partial/\partial q.
\]
In spite of the presence of this additional nonlocal symmetry it is rather
unlikely that the case of $\nu=1$ is integrable but
the definitive settling of this issue requires further research.

Recall now that a {\em local conservation law} for the system
(\ref{Vladimirov:mainspat}) is a relation of the form
\[
D_t R-D_x S,
\]
vanishing modulo (\ref{Vladimirov:mainspat}) and its differential consequences,
where $R,\,S$ depend on $x,t,u,\rho$ and
a finite number of $x$-derivatives of $u$ and $\rho$,  $D_t$ and
$D_x$ denote the total derivative w.r.t.\ the temporal and spatial
variables, respectively, cf.~\cite{Ibragimov,Olver,BlumanCheviakov}
for details.

The local conservation laws associated with
(\ref{Vladimirov:mainspat}) read as follows:
\begin{equation}\label{dens1}
\hspace*{-5mm}\begin{aligned} R_1 &= u,\quad & S_1
&=-\frac{\beta}{\nu+2}\rho^{\nu+2}-\gamma\left(\rho^{\nu+1} \rho_{x}
\right)_x\\
R_2 &=1/\rho,\quad & S_2 &=u,\\
R_3 &=u \,\sin(\omega \,x),\quad & S_3 &=
\displaystyle\frac{\cos(\omega \,x)}{\omega}\,\beta
\rho^{\nu+1}\rho_x-\sin(\omega x)\gamma \left(\rho^{\nu+1}
\rho_{x}\right)_x,\\
R_4 &=u \,\cos(\omega x) ,\quad & S_4 &=
-\displaystyle\frac{\sin(\omega x)}{\omega}\beta
\rho^{\nu+1}\rho_x-\cos(\omega x)\gamma \left(\rho^{\nu+1}
\rho_{x}\right)_x,\\
R_5 &=t u+x/\rho,\quad & S_5 &= t S_1+ x u,\\
R_6 &=t \,u\, \sin(\omega x)- \displaystyle\frac{\cos(\omega
x)}{\omega\rho},
 \quad & S_6 &= t S_3-u\,\frac{\cos(\omega x)}{\omega},\\
R_7 &=t \,u \cos(\omega x)+\displaystyle\frac{\sin(\omega
x)}{\omega\rho}, \quad & S_7 &= t \,S_4+ u\,\frac{\sin(\omega
x)}{\omega},
\end{aligned}
\end{equation}
where $\omega=(\beta/\gamma)^{1/2}$, and no further conserved
quantities with local densities for the generic values of parameters
in (\ref{Vladimirov:mainspat}) seem to exist. We intend to address this
issue in more detail in the future work.

Note that if $\beta/\gamma<0$, then some of the above $R_i$ and $S_i$
become complex, and it is convenient to use a slightly different
basis of conservation laws instead of (\ref{dens1}), namely
\begin{equation}\label{dens1a}
\hspace*{-5mm}\begin{aligned} \tilde R_1 &= u,\quad & \tilde S_1
&=-\frac{\beta}{\nu+2}\rho^{\nu+2}-\gamma\left(\rho^{\nu+1} \rho_{x}
\right)_x\\
\tilde R_2 &=1/\rho,\quad & \tilde S_2 &=u,\\
\tilde R_3 &=u \,\sinh(\tilde\omega \,x),\quad & \tilde S_3 &=
\displaystyle-\frac{\cosh(\tilde\omega \,x)}{\tilde\omega}\,\beta
\rho^{\nu+1}\rho_x-\sinh(\tilde\omega x)\gamma \left(\rho^{\nu+1}
\rho_{x}\right)_x,\\
\tilde R_4 &=u \,\cosh(\tilde\omega x) ,\quad & \tilde S_4 &=
-\displaystyle\frac{\sinh(\tilde\omega x)}{\tilde\omega}\beta
\rho^{\nu+1}\rho_x-\cosh(\tilde\omega x)\gamma \left(\rho^{\nu+1}
\rho_{x}\right)_x,\\
\tilde R_5 &=t u+x/\rho,\quad & \tilde S_5 &= t \tilde S_1+ x u,\\
\tilde R_6 &=t \,u\, \sinh(\tilde\omega x)+ \displaystyle\frac{\cosh(\tilde\omega
x)}{\tilde\omega\rho},
 \quad & \tilde S_6 &= t \tilde S_3+u\,\frac{\cosh(\tilde\omega x)}{\tilde\omega},\\
\tilde R_7 &=t \,u \cosh(\tilde\omega x)+\displaystyle\frac{\sinh(\tilde\omega
x)}{\tilde\omega\rho}, \quad & \tilde S_7 &= t \tilde S_4+ u\frac{\sinh(\tilde\omega
x)}{\tilde\omega},
\end{aligned}
\end{equation}
where $\tilde\omega=(-\beta/\gamma)^{1/2}$. It is immediate that for $\beta/\gamma<0$
the conservation laws with the densities and fluxes (\ref{dens1a}) are real functions.

\section{Qualitative study of system of ODEs describing the
traveling wave solutions for (\ref{Vladimirov:mainspat})}

We are going to analyze a set of traveling wave (TW) solutions
having the form \begin{equation}\label{Vladimirov:TWsol}
u(t,\,x)=U(z), \qquad \rho(t,\,x)=R(z), \qquad z=x-s\,t,
\end{equation} where $s$ is the velocity of TW.
 Inserting the ansatz
(\ref{Vladimirov:TWsol}) into the second equation of the system
(\ref{Vladimirov:mainspat}) we immediately get the quadrature
\begin{equation}\label{Vladimirov:firstint}
U(z)=C_1-\frac{s}{R(z)},
\end{equation}
where $C_1$ is the integration constant. In what follows we
assume that $C_1=s/R_1$, where $0<R_1=const$. Such a choice leads to
the following asymptotic behavior:\[ \lim\limits_{|z|\,\to\,
\infty}u(t,\,x)=0, \qquad \lim\limits_{|z|\,\to\,
\infty}\rho(t,\,x)=R_1.\]

Inserting the ansatz (\ref{Vladimirov:TWsol}) into the first
equation of the system (\ref{Vladimirov:mainspat}), and using the
equation (\ref{Vladimirov:firstint}), we obtain, after one
integration, the following ODE:
\begin{equation}\label{Vladimirov:ges}
\frac{s^{2}}{R}+\frac{\beta}{\nu+2} R^{\nu+2}+
\gamma\left[R^{\nu+1}\frac
{d^2\,R}{d\,z^2}+(\nu+1)R^{\nu}\left[\frac{d\,R}{d\,z}\right]^{2}\right]=E,
\end{equation}
where
\begin{equation}\label{Vladimirov:const_e}
E=\frac{s^{2}}{R_{1}}+\frac{\beta}{\nu+2} R_{1}^{\nu+2},
\end{equation}
is a constant of integration, defined by the conditions on $+\infty$.

Let us write equation (\ref{Vladimirov:ges})  in the form of the first order dynamic system:
\begin{equation}\label{Vladimirov:ds1}
\left\{
\begin{array}{ll}
\frac{dR}{d\,z}=Y \\ \\
\frac{dY}{d\,z}=
\left(\gamma R^{\nu+2}\right)^{-1}
\left\{ER-\left[s^{2}+\frac{\beta}{\nu+2}R^{\nu+3}+\gamma(\nu+1)R^{\nu+1}Y^{2}\right]\right\}.
\end{array} \right.
\end{equation}
%8
 It is evident, that all isolated stationary points of the system (\ref{Vladimirov:ds1}) are located on the horizontal axis  $OR$.  They  are determined by  solutions of   the algebraic equation
\begin{equation}\label{Vladimirov:cp_spat}
P(R)=\frac{\beta}{\nu+2}R^{\nu+3}-ER+s^{2}=0.
\end{equation}
As can be easily seen,  one of the roots of equation (\ref{Vladimirov:cp_spat}) coincides with $R_{1}$.
Location of the second real positive root, $R_2$, depends on relations between the parameters.
It is placed to the right from $R_1$  if
 $s^2$ satisfies inequality
\begin{equation}\label{Vladimirov:dlowerb}
s^{2}>s_{1}^{2}=\beta R^{\nu+3}_{1},
\end{equation}
and to the left of it if the $s^2<s_1^2$. It can also be shown that  for $ \nu$ being natural number or zero the polynomial $P(R)$ has the  representation
\begin{equation}\label{Vladimirov:p_repr}
P(R)=(R-R_{1})(R-R_{2})\Psi (R),
\end{equation}
%9
where
\[
\begin{array}{llc}
\Psi(R)&=&\frac{\beta}{(\nu+2)\left(R_{2}-R_{1}\right)}\{R^{\nu+1}(R_{2}-R_{1}) +R^{\nu}(R_{2}^2-R_{1}^2)+\dots
+(R_{2}^{\nu+2}-R_{1}^{\nu+2}) \}.
\end{array}
\]
Note that $\Psi(R)$ is positive, when $R>0$.

Analysis of system's (\ref{Vladimirov:ds1}) linearization matrix
\begin{equation}\label{Vladimirov:mlins}
\hat M(R_i,\,\,0)=\left[
\begin{array}{cc}
0 & 1 \\
\left(\gamma\,R_i^{\nu+2} \right)^{-1}\Psi(R_i)(R_j-R_i) & 0
\end{array}
\right],
\qquad i=1,2, \qquad j\ne i
\end{equation}
shows, that the stationary points $(R_{1},0)$ is a saddle, while the point  $(R_{2},0)$ is a center if either $\gamma>0$ and $|s|>s_1$, or $\gamma<0$ and $|s|<s_1$.   In both of these cases the system (\ref{Vladimirov:ds1}) has only such stationary points, which are characteristic to the Hamiltonian systems. It suggests that, by a proper choice of the integrating factor, (\ref{Vladimirov:ds1}) can be rewritten in the Hamiltonian form. In fact, introducing the new independent variable $\frac{d}{d\,T}=2 \gamma R^{\nu+2} \frac{d}{d\,z}$ we get:
 \[
 \frac{d\,R}{d\,T}=\frac{\partial\,H}{\partial\,Y}, \qquad \frac{d\,Y}{d\,T}=-\frac{\partial\,H}{\partial\,R}, \qquad
 \]
 with
\begin{equation}\label{Vladimirov:hfun1}
 H=2s^{2}\frac{R^{\nu+1}}{\nu+1}+\frac{\beta}{(\nu+2)^{2}}R^{2(\nu+2)}+\gamma Y^{2}R^{2(\nu+1)}-2E\frac{R^{\nu+2}}{\nu+2}.
\end{equation}

The existence of the Hamiltonian function implies that the stationary point $(R_2,\,0)$ keeps to be surrounded
by the one-parameter family of periodic trajectories after the nonlinear terms {are} added. Now we are interested on whether or not this set is bounded or unbounded. In the first case its natural limit is a homoclinic trajectory bi-asymptotic to the saddle $(R_1,\,0)$. The homoclinic trajectory, in turn, corresponds to the solitary wave solution of the source system (\ref{Vladimirov:mainspat}).

To answer the above question, we analyze the Hamiltonian function (\ref{Vladimirov:hfun1}) which remains constant on the trajectories of the dynamical system. Thus, the equation for saddle sepratrices takes the form
 \begin{equation}\label{Vladimirov:separ1}
Y=\pm \frac{\sqrt{Q(R)} }{\sqrt{|\gamma|}R^{\nu+1}},\end{equation}
where
\begin{equation}\label{Eq:forQ}
 Q(R)=\frac{\gamma}{|\gamma|}\,\left\{H_1+2\,E\,\frac{R^{\nu+2}}{\nu+2}-\left[ 2 s^2 \frac{R^{\nu+1}}{\nu+1}+
   \frac{\beta}{(\nu+2)^2}\,R^{2(\nu+2)}\right]  \right\},
\end{equation}
and
\begin{equation}\label{Ham1}
H_1=H(R_1,\,0)=\frac{2\,R_1^{\nu+1}}{2\,(\nu+2)}\left\{s^2-  \frac{\beta(\nu+1)}{2\,(\nu+2)}R_1^{\nu+3} \right\}.
\end{equation}
Since the incoming and outgoing separatrices are symmetrical with respect to  $OR$ axis, it is sufficient to consider the upper separatrix $Y_{+}$ lying, depending on the sign of $\gamma$, to the right or to the left of the saddle point $(R_1,\,0)$.

First, let us note, that
\[
Q^\prime (R)|_{R=R_1}
=-2\frac{\gamma}{|\gamma |} R^\nu\,P(R)|_{R=R_1}=0,
\]
while
\[
Q^{\prime\prime} (R)|_{R=R_1}=2\frac{\gamma}{|\gamma |},R_1^{\nu-1}\left[s^2-\beta\, R_1^{\nu+3}  \right].
\]
Thus $Q(R)$ has a local minimum in $R_1$ when either
\begin{equation}
\label{smore}
s>s_1  \quad \mathrm{and} \quad\gamma >0,
\end{equation}
or
\begin{equation}
\label{sless}
s<s_1  \quad \mathrm{and} \quad\gamma <0.
\end{equation}
Let us observe that
$H_1$ is positive whenever
$$
s^2-  \frac{\beta(\nu+1)}{2\,(\nu+2)}R_1^{\nu+3}>0.
$$

Now we can  concentrate on the case (\ref{smore}) for which we are interested in the behaviour of $Y_{+}(R)$ to the right of $(R_1,\,0)$. The function $Y_{+}(R)$ is a growing function when  $R-R_1>0$ is small, but since the coefficient of the highest power of $R$ in the function $Q(R)$ is negative, then for greater $R$ it becomes a decreasing function, intersecting the horizontal axis $OR$ at least once.  The coordinate of the first intersection is denoted as $R_{*}^{+}$. It is evident that $R_{*}^{+}>R_1$, and
\[
\lim_{R\to R_{*}^{+}-0}Q(R)=+0.
\]
Furthermore, the function
$
Q'(R)=-2 R^\nu P(R)
$
is positive when $R<R_2$ and negative when $R>R_2$. Let us consider the expression
\begin{equation}\label{Ypr}
\frac{d\,Y_{+}(R)}{d\,R}=\frac{R Q'(R)- 2 (\nu+1) Q(R)}{2 \sqrt{|\gamma| Q(R)} R^{{\nu+2}}}=
-\frac{\left({\gamma}/{|\gamma|}\right)\,R^{\nu+1}P(R)+(\nu+1)\,Q(R)}{\sqrt{|\gamma|\, Q(R)} R^{{\nu+2}}}.
\end{equation}
If $R_{*}^{+}>R_2$, then this expression tends to $-\infty$ as $R\to R_{*}^{+}-0$, otherwise it tends to  $+\infty$.
But the last option is impossible since it contradicts the assertion that $R_{*}^{+}$ is the point of the first intersection of $Q(R)$ with the horizontal axis. Form this we conclude that  $R_1<R_2<R_{*}^{+}$ and the curves $Y_{\pm}(R)$ form a homoclinic loop.

Now let us assume that the conditions (\ref{sless}) take place. Obviously we are interested in the behavior of the left upper branch of the separatrice $Y(R)$ in this case. As it follows from (\ref{Eq:forQ}),
$
\lim\limits_{R\to +0} Q(R)=-H(R_1)<0,
$
thus $Q(R)$, which initially grows with the growth of  $0<R_1-R<<1$, should intersect the horizontal axis in some point
$R_{*}^{-}$ such that $0<R_{*}^{-}<R_1.$
The analysis of the formula (\ref{Ypr}) suggests that if $0<R_{*}^{-}<R_2$ then ${d\,Y_+(R)}/{d\,R}$ tends to $+\infty$ as $R$ tends to $R_{*}^{-}+0$ and to $-\infty$ if $R_{*}^{-}>R_2$. The last assumption contradicts the assertion that $R_{*}^{-}$ is the point in which $Q(R)$ intersects the horizontal axis. Hence the pair $Y_{\pm}(R)$, under the above conditions, the homoclinic loop lying to the left of $(R_1,\,0)$ and surrounding an open set $\Omega \,\ni\,(R_2,\,0)$ filled with the periodic trajectories.

The result obtained can be formulated in the following:
\vspace{2mm}

\begin{thm} Let us assume that  $\nu$ is a natural number or zero.
The system (\ref{Vladimirov:ds1}) possesses the homoclinic solutions corresponding to the solitary wave
solutions of the source system (\ref{Vladimirov:mainspat}) and satisfying the initial conditions
$$\lim\limits_{|z|\to \infty}U(z)=0, \quad \lim\limits_{|z|\to \infty}R=R_1>0$$
if either $\gamma<0$ and the  inequalities
\begin{equation}\label{gammin}
\frac{\beta\,(\nu+1)}{2\,(\nu+2)}\,R_1^{\nu+3}<\,s^2<\beta\,R_1^{\nu+3},
\end{equation}
hold, or $\gamma>0$ and
\begin{equation}\label{gamplus}
s^2>\beta\,R_1^{\nu+3}.
\end{equation}
In the first case the system (\ref{Vladimirov:mainspat}) possesses a one-parameter family of the soliton-like solution
that describes the waves of rarefaction,
while in the second case - the family of the soliton-like solutions
that describes the waves of compression.

\end{thm}

\begin{figure}[h]
  \centering
  {\includegraphics[totalheight=1.42 in]{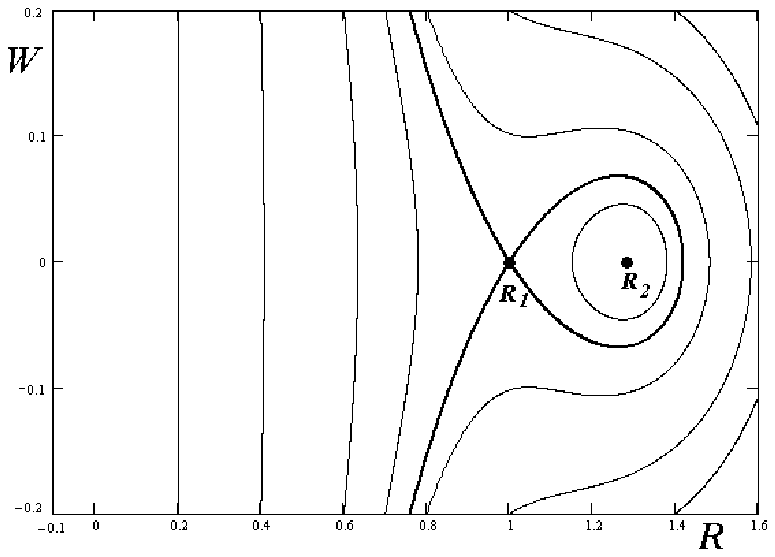}} \quad
  {\includegraphics[totalheight=1.42 in]{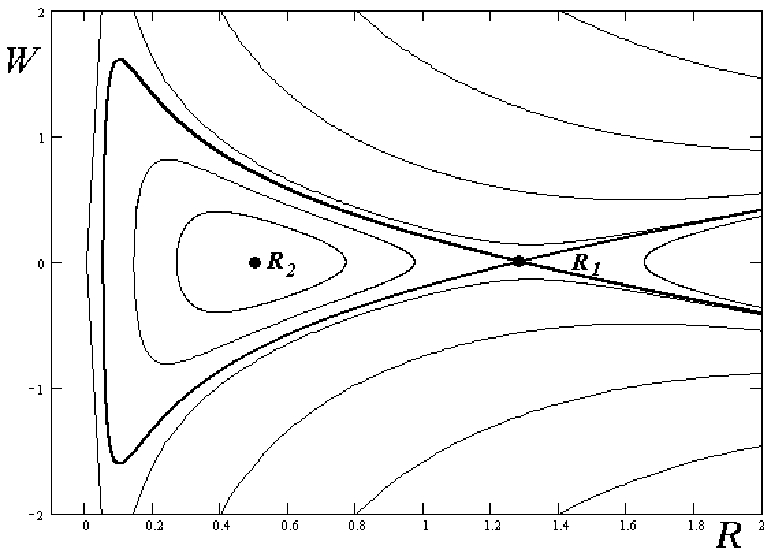}} \quad
  {\includegraphics[totalheight=1.42 in]{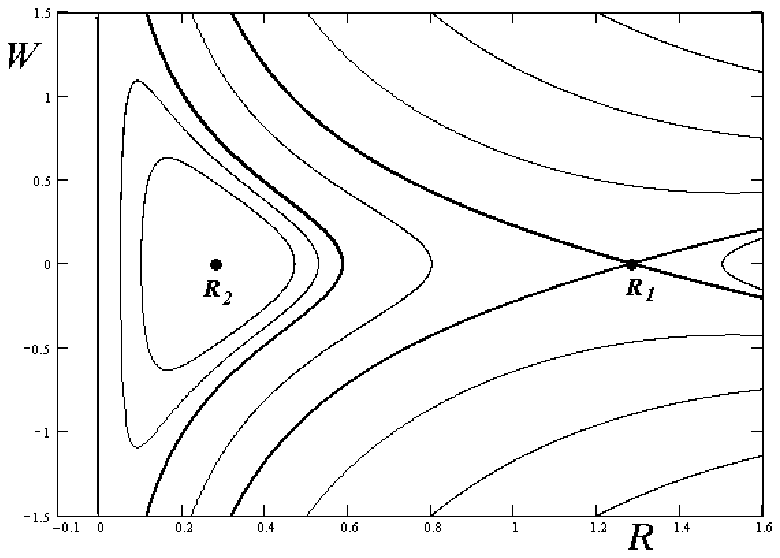}}
  \caption{ Phase portraits of the dynamic system (\ref{Vladimirov:ds1}) obtained
for  $\nu=0$ and $\beta=1.75$:  a) $\gamma=3,\,\,s=1.6$, $R_1=1$;
b) $\gamma=-3$, $s=1.1$, $R_1=1.282$;  and c)
$\gamma=-3$, $s=0.7$, $R_1=1.282$. }\label{Fig:ris1}
\end{figure}

Below we give the typical phase portraits of the system
(\ref{Vladimirov:ds1}). To be specific, the plots were made for
 $\beta=1.75$ and $\nu=0$. The remaining parameters
vary from case to case. The phase portrait corresponding to
$\gamma=3$, $s=1.6$ and $R_1=1$ is shown in
Fig.~\ref{Fig:ris1}~a. The phase portrait of the system contains a
closed loop which corresponds to the wave of compression.

%\begin{figure}[h]
  % Requires \usepackage{graphicx}
%\includegraphics[totalheight=1.35 in]{phas_a}
%\includegraphics[totalheight=1.35 in]{phas_b_s1}
%\includegraphics[totalheight=1.35 in]{phas_c}
%\centerline{  a \hspace{3. cm} b \hspace{5cm} c} \caption{ Phase
%portraits of the dynamic system (\ref{Vladimirov:ds1}) obtained for
% $\nu=0$, and $\beta=1.75$:  a) $\gamma=3,\,\,s=1.6$, $R_1=1$; b)
%$\gamma=-3,\,\,s=1.6$, $R_1=1.282$;  and c) $\gamma=-3,\,\,s=0.7$,
%$R_1=1.282$. }\label{Fig:ris1}
%\end{figure}

Fig.~\ref{Fig:ris1}~b, plotted for   $\gamma=-3$,  $s=1.1$ and
$R_1=1.282$, contains the closed loop directed towards the vertical
axis. This loop corresponds to the solitary wave of rarefaction.
Fig.~\ref{Fig:ris1}~c plotted for   $\gamma=-3$, $R_1=1.282$ and
$s=0.7$ does not contain the closed loop since the inequalities
(\ref{gammin}) are not satisfied.

\section{Spectral stability of the stationary solutions }

In the study of spectral stability of TW solutions, it is
helpful to pass to new independent variables
\[
\bar{t}=t, \qquad \bar{z}=x-s\,t,
\]
in which the invariant TW solutions (\ref{Vladimirov:TWsol}) become
stationary. Since the main part of the analysis is performed numerically,
we confine ourselves to the case $\nu=0.$ In the
new variables the system (\ref{Vladimirov:mainspat}) reads as
follows:
\begin{equation}\label{Vladimirov:newspat}
\left\{
\begin{array}{ll}
u_{\bar t}-s\,u_{\bar z}+\beta\rho\,\rho_{\bar
z}+\gamma[\rho\rho_{\bar z \bar z \bar z}+
3\,\rho_{\bar z}\rho_{\bar z\bar z}]=0,\\
\rho_{\bar t}-s\,\rho_{\bar z}+\rho^{2}u_{\bar z} =0,\end{array}
\right.
\end{equation}
 (for the sake of simplicity, the bars will be omitted from now on).
We restrict ourselves to the analysis of spectral stability
\cite{Evans3,Evans4,Sandstede} of the TW solution $\left(U(z),R(z)\right)$, and consider the perturbations of the following form:
\begin{equation}\label{Vladimirov:stab_ans}
u(t,\,z)=U(z)+\epsilon\,\exp{[\lambda\,t]}\,f(z), \qquad
\rho(t,\,z)=R(z)+\epsilon\,\exp{[\lambda\,t]}\,g(z),\end{equation}
where $\lambda$ is the spectral parameter, and  $|\epsilon|\ll\, 1.$

Inserting the ansatz (\ref{Vladimirov:stab_ans}) into the system
(\ref{Vladimirov:newspat}) and neglecting the $O\left(|\epsilon|^2
\right)$ terms, we obtain the system linearized about the travelling wave solutions:
\begin{eqnarray}\begin{array}{l}
f\lambda  - s f^\prime  + R \gamma g^{\prime\prime\prime}  + \beta g
R^\prime  +
  3 \gamma g^{\prime \prime} R^\prime + g^\prime (\beta R + 3 \gamma R^{\prime\prime}) +
  g \gamma R^{\prime\prime\prime} =0, \vspace{0.1cm} \\
R^2 f^\prime - s g^\prime + g(\lambda + 2 R U^\prime)=0,
\end{array}
\qquad\label{Vladimir:var2}
\end{eqnarray}where  the prime denotes
the derivative with respect to $z$.

Rewrite (\ref{Vladimir:var2}) as a first-order dynamical system
\begin{equation}\label{dyn_syst}
Y^\prime=A Y,
\end{equation}
 where $Y=\left(g,\,\eta,\,\chi,\,f  \right)^{tr}$,
 $$A=\left(\begin{array}{cccc}
 0 & 1 & 0 & 0 \\
0&0&1&0 \\
a_1&a_2&a_3&a_4 \\
a_5&a_6&0&0
\end{array}
 \right)$$
$\displaystyle  a_1=-\frac{\beta R^\prime+\gamma
R^{\prime\prime\prime}+R^{-2}s\left(\lambda+2RU^\prime\right)}{\gamma
R}$,
 $\displaystyle  a_2=\frac{s^2-\beta R^3 -3R^2\gamma
R^{\prime\prime}}{\gamma R^3}$,
 $\displaystyle  a_3=-\frac{3R^\prime}{R}$,
 $\displaystyle  a_4=-\frac{\lambda}{R\gamma}$,  $\displaystyle  a_5=-\frac{\lambda +2RU^\prime}{R^2}$,
  $\displaystyle  a_6=\frac{s}{R^2}$.
Since $R(z)$ and $U(z)$  tend to their limiting values ($R_1$ and
$0$ respectively) as $|z|$ increases, the dynamical system under
study asymptotically tends to the system with constant
coefficients
 \begin{equation}\label{Vladimir:varoninf}
Y^\prime=A_{\infty} Y,
\end{equation}
where
$$  A_\infty=\left(\begin{array}{cccc}
 0 & 1 & 0 & 0 \\
0&0&1&0 \\
\displaystyle -\frac{s\lambda}{\gamma R_1^3} & \displaystyle
\frac{s^2-\beta R_1^3}{\gamma R_1^3} &
0 & \displaystyle -\frac{\lambda}{\gamma R_1} \\
\displaystyle -\frac{\lambda}{R_1^2} & \displaystyle \frac{s}{R_1^2}
& 0&0
\end{array}
 \right)$$

It is obvious that the linearized system (\ref{Vladimir:var2}) can
be treated as a spectral problem

 \begin{equation}\label{Vladimir:Vareqvec} Ly=\lambda
y, \qquad  y=\left(f,\,g\right)^{tr},\end{equation} for the operator
$$
L= \left(
\begin{array}{cc}
- s \partial _z & R \gamma \partial_{z\,z\,z} +
  3 \gamma  R^\prime  \partial_{z\, z}+ (\beta R + 3 \gamma R^{\prime\prime})\partial_z  +
 \beta R^\prime + \gamma R^{\prime\prime\prime}  \vspace{0.1cm} \\
R^2 \partial_z  &   2 R U^\prime - s \partial_z
\end{array}
\right).
$$
Recall that the set of all possible values of
$\lambda\,\in\,\mathbb{C}$ for which the equation
\[
\frac{\partial}{\partial\,t}\left[\begin{array}{l} \bar{u}(t,\,z) \\ \bar\rho(t,\,z) \end{array}\right]=L\,\left[\begin{array}{l} \bar{u}(t,\,z) \\ \bar\rho(t,\,z) \end{array}\right]
\]
has  nontrivial solutions of the form $\exp{[\lambda\,t]} \left[ f(z),\,g(z)\right]^{tr}$ is called the spectrum $\sigma$ of the operator $L$.  The homoclinic  solution
$\left[U(z),\,R(z)\right]^{tr}$ is said to be spectrally stable if no
possible eigenvalue $\lambda$ belongs to the right half-plane of the
complex plane.

\begin{rmk} It follows from the translation invariance
of the system (\ref{Vladimirov:newspat})
 that zero belongs to the spectrum of $L$.
\end{rmk}

As usually, we distinguish the essential spectrum
$\sigma_{ess}\,\subset\,\sigma$, and the discrete spectrum
$\sigma_{discr}\,\subset\,\sigma$. Being somewhat informal, we can
treat $\sigma_{ess}$ and $\sigma_{discr}$ as the subsets
responsible, respectively,  for the stability of the stationary
solutions $\left(0,\,R_1 \right)$, and the solution $\left[
U(z),\,R(z) \right]^{tr}$ itself.

Now we are going to state the conditions which guarantee that the set
$\sigma_{cont}\,\bigcap\,\mathbb{C}^{+}=\emptyset$. In the limiting case  $|z|
\rightarrow\,\pm \,\infty$, the variational system turns into a
linear system with constant coefficients,
\begin{eqnarray}
-\gamma\, R_1\,g^{\prime\prime\prime}
+s\,f^\prime-\beta\,R_1\,g^\prime=\lambda f,\label{Vladimir:varlim1}\\
s\,g^\prime-R_1^2\,f^\prime=\lambda\,g,\qquad\nonumber\end{eqnarray}
%We also assume that the eigenvectors
%$\left[f(z),\,g(z)\right]$ belong to the space of tempered
%distributions $\mathcal{S'(R)}$ \cite{Maurin}. With this assumption
Location of the essential spectrum can be determined using the
Fourier transform. Applying the latter to the
system (\ref{Vladimir:varlim1}) yields
\begin{equation}
 \hat M(\xi,\,\lambda)\,\left( \begin{array}{c} \hat f(\xi) \\ \hat g(\xi)
\end{array}\right)= \left( \begin{array}{ll} \lambda +i\,\xi \,s, &
i\,\xi\,R_1\,(\gamma\,\xi^2-\beta) \\
-i\,\xi\,R_1^2, & \lambda+i\,\xi\,s
\end{array}\right)\left( \begin{array}{c} \hat f(\xi) \\ \hat g(\xi)
\end{array}\right)=0,
\end{equation}
where $\hat f(\xi)$ and $\hat g(\xi)$ are respectively the Fourier transforms of
$f(z)$ and $g(z)$. Equating the determinant of the
matrix $\hat M(\xi,\,\lambda)$ to zero, we obtain the expression
for eigenvalues:
\begin{equation}\label{eignvcont}
\lambda_{1,\,2}={-\,i\,\xi\,s\,\pm\sqrt{\left(\gamma\,\xi^2-\beta\right)\xi^2\,R_1^{3}}},
\quad \xi\,\in\,R.
\end{equation}
Thus, the following assertion holds.

\begin{st}  {  If $\gamma<0$ and $\beta>0$, then
the essential spectrum $\sigma_{ess}$ does not intersect the positive half-plane
$\mathbb{C}^+$. On the other hand, if both $\beta$ and $\gamma$ are positive, then $\sigma_{ess}$   has a
non-empty intersection with $\mathbb{C}^+$ and the stationary
solution $(0,\,R_1)$  is unstable.}
\end{st} %\vspace{2mm}

\section{ Numerical study of the discrete spectrum and
the dynamical behavior of solitary wave solutions}

An efficient tool for the  study of discrete spectrum of a
linearized operator is provided by the so called Evans function
$E(\lambda)$, which is an analytic function of the spectral
parameter $\lambda$. The zeroes of $E(\lambda)$ correspond to the
eigenvalues of the linearized operators that belong to the discrete
spectrum \cite{Evans3,Sandstede}.

The Evans function is constructed by evolving the linearized system,
depending on $\lambda$, starting from the points of initiation lying
at $-\infty$ in the unstable invariant manifold, and from the points
at $+\infty$ lying in the stable one. The solutions (mostly
extrapolated numerically) are then calculated for some fixed value
of $z_0$ (usually for $z_0=0$), and the value of the Wronskian at this
point determines $E(\lambda)$. If for some $\lambda_0$ the Evans
function vanishes, then the intersection of the stable and unstable
manifolds is nontrivial, there exists the corresponding eigenvector which
belongs to the Hilbert space of square integrable functions
$L^2(R)$ and, thus, $\lambda_0\,\in\,\sigma_{discr}.$

\begin{center}
\begin{figure}[h]
  % Requires \usepackage{graphicx}
  \includegraphics[totalheight=2.5in]{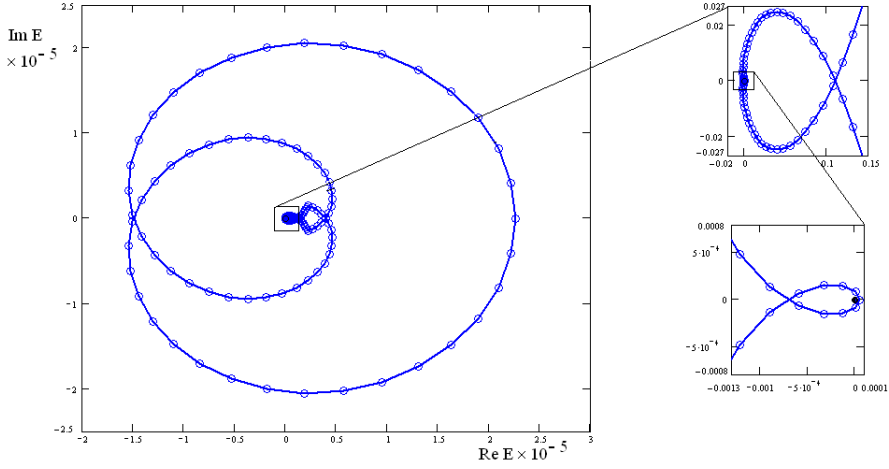}
 \caption{The real versus imaginary part of $E(\lambda)$ for
    $s=1.6$,
$\beta=1.75$,  ${\gamma=-3}$,  $R_1=1.282$. The spectral parameter
varies along the border of the half-circle with the radius $b=12$,
symmetric w.r.t.\ the horizontal axis and separated from the vertical
axis by a small offset $a=0.03$. }\label{Fig:Nyq1}
\end{figure}
\end{center}

The analyticity of the Evans function  enables to calculate the number of roots of the equation $E(\lambda)=0$,   together with their multiplicities, within the compact domain $B\subset\mathbb{C}^+$, using the well-known formula (see e.g. \cite{Maurin},~Ch.~XV )
\[
N=\sum_{j=1}^{n_0} k_j=\frac{1}{2\,\pi\,i}\,\oint_{\partial\,B} \frac{E^\prime(\lambda)}{E(\lambda)}d\,\lambda,
\]
where $n_0$ is the number of roots, while $k_j$ is the multiplicity of the $j$-th root. The number $N$, called the {\it winding number} \cite{Evans4,derks,Gottwald,Blank}, determines how many turns makes the vector $\left(Re{E(\lambda)},\,Im{E(\lambda)} \right)$ around the origin as $\lambda$ runs along the contour $\partial\,B$.
 Since we are unable to
integrate numerically over an unbounded region, it is  necessary to
become convinced that the Evans function is nonzero for large
$|\lambda|$ belonging to the positive half-plane of the complex
plane (some relevant estimates  are presented in Appendix 1).
Thus, calculating $N$ for sufficiently large $B$ (usually $B$ is a
semicircle lying in $\mathbb{C}^+$) and analyzing the behavior of
$E(\lambda)$ for large $|\lambda|$ can give a hint regarding the
location of $\sigma_{discr}.$

\begin{center}
\begin{figure}[h]
  % Requires \usepackage{graphicx}
  \includegraphics[totalheight=2.5in]{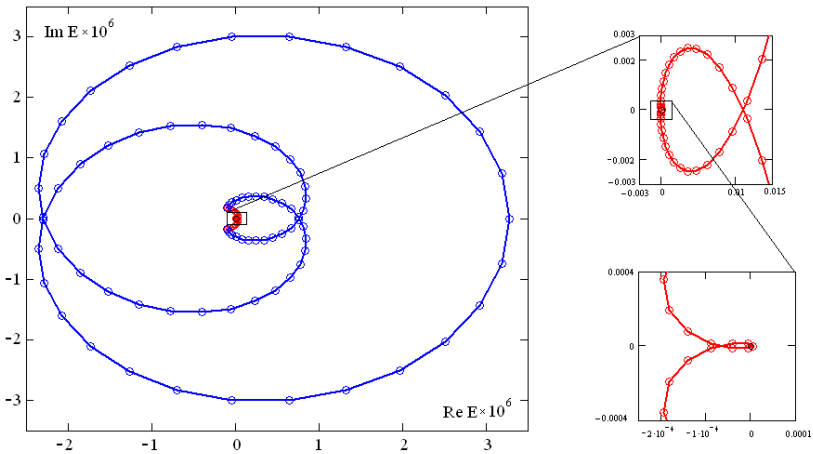}
 \caption{The real versus imaginary part of $E(\lambda)$ for
    $s=1.6$,
$\beta=1.75$,  ${\gamma=-3}$,  $R_1=1.282$. The spectral parameter
varies along the border of the half-circle with the radius $b=20$,
symmetric w.r.t.\ the horizontal axis and separated from the vertical
axis by a small offset $a=0.03$. }\label{Fig:Nyq2}
\end{figure}
\end{center}

The construction of the Evans function is performed as follows. For
large $z$ the spectral problem becomes close to the equation
(\ref{Vladimir:varoninf}). Solutions to this equation, which are
easily calculated, form at $+\infty$ a stable manifold $U^+$
spanned by $k$ independent eigenvectors
$\left\{\vec{v_j}^+\right\}_{j=1}^k$ of the matrix $A_\infty$,
corresponding to the eigenvalues $\lambda_j$ with  the negative real
part. At $-\infty$ solutions of the equation
(\ref{Vladimir:varoninf}) form an unstable manifold $U^-$ spanned
by $m$ independent eigenvectors $\left\{\vec{v_j}^-\right\}_{j=1}^m$
of the matrix $A_\infty$ corresponding to the eigenvalues with the
positive real parts. The construction of the Evans function becomes
possible when these two sets of vectors are complementary, i.e., when
$m=n-k$.

So, initializing (\ref{dyn_syst}) with $k$ vectors $\vec{v_j}^+$
from  $U^+$ and with $n-k$ independent vectors $\vec{v_j}^-$ from
$U^-$ and solving the system towards $z=0$, we obtain two sets of
vectors:
$$V^-~=~\left\{V_1^-(0,\,\lambda),\, V_2^-(0,\,\lambda),\dots,
V_{n-k}^-(0,\,\lambda) \right\}$$ and
$$V^+=\left\{V_{1}^+(0,\,\lambda),\, V_{2}^+(0,\,\lambda),\dots,
V_k^+(0,\,\lambda)\right\},$$ being the analytic functions of $\lambda$.
These sets have nontrivial intersection if and only if  $\lambda$
belongs to the discrete spectrum $\sigma_{discr}$. Therefore the
analysis of zeroes of the Evans function defined as
\begin{equation}\label{Vladimir:Evansdf}
E[\lambda]=\det \left[V_1^-(0,\,\lambda),\, V_2^-(0,\,\lambda),\dots,
V_{n-k}^-(0,\,\lambda),\, V_{1}^+(0,\,\lambda),\,
V_{2}^+(0,\,\lambda),\dots, V_k^+(0,\,\lambda)   \right]
\end{equation}
reveals the location of discrete spectrum of the operator $L$.

We perform the construction of the Evans function based on a
numerical procedure.  An appropriate method depends on the
dimensions of invariant manifolds $U^+$ and $U^-$. In our case
both stable and unstable invariant manifolds of the matrix
$A_{\infty}$ happen to be two-dimensional. Prolongation of $U^{\pm}$
in multi-dimensional cases encounters the well-known obstacles
\cite{briges} which can be overcome by employing the
exterior algebra.

Using the wedge product we derive a $k$-form in the vector
space $\wedge^k (\mathbb{C}^n)$ built from the basis elements  of
the vector space $\mathbb{C}^n$. In our case $n=4$ (see Appendix 2),
and the two-forms belonging to the space $\wedge^2 (\mathbb{C}^4)$
correspond to the invariant manifolds $U^{+}$ and $U^{-}$.
As the basis we choose the vectors  $w_1=e_1 \wedge e_2$,
$w_2=e_1 \wedge e_3$, $w_3=e_1 \wedge e_4$, $w_4=e_2 \wedge e_3$,
$w_5=e_2 \wedge e_4$, $w_6=e_3 \wedge e_4$. Mapping the dynamical
system (\ref{dyn_syst}) into the space spanned by $w_1,...w_6$,  we obtain:
\begin{equation}\label{extend}
U^\prime =A^{(2)}U,
\end{equation}
where
$$
A^{(2)}=\left(
\begin{array}{cccccc}
0 & 1  & 0 & 0 & 0 & 0 \\
a_2 & a_3 & a_4 & 1 & 0 & 0 \\
a_6 & 0 & 0 & 0 & 1 & 0\\
-a_1 & 0 & 0 & a_3 & a_4 & 0 \\
-a_5 & 0 & 0 &  0 & 0 & 1 \\
0 & -a_5 & a_1 & -a_6 & a_2 & a_3
\end{array} \right)
$$
The set of  eigenvalues of the matrix  $A^{(2)}$ consists of all
possible sums of eigenvalues of the matrix $A$. Therefore the
manifolds $U^{\pm}$ are the solutions of the system
 (\ref{extend}), for which
$$\lim_{z \rightarrow \pm \infty}
e^{-\mu _{\pm} z}U^{\pm} = \hat v^{\pm},$$ where $\hat v^{+}(\hat v^{-})$ is an
eigenvector of the matrix $A^{(2)}_\infty$ corresponding to the
eigenvalue $\mu_+(\mu_-)$ with the smallest negative (largest
positive) real part. The Evans function can be represented in terms of
solutions $U^{\pm}$  in the following fashion  \cite{Gottwald}:
\begin{equation}\label{thisexpr}E(\lambda)=\exp\left\{-\int_0^z
\mbox{Tr}(A^{(2)})dz\right\}U^+ (z,\lambda) \wedge U^-
(z,\lambda).\end{equation}
In  evaluating the expression
(\ref{thisexpr}) we employed the relation
$$U^+ (z,\lambda) \wedge U^-
(z,\lambda)=\left\langle U^-, \Sigma U^+\right\rangle_R,$$ where
$\langle \cdot, \cdot \rangle$ is the scalar product in the space $R^6$,
$$\Sigma=\left(
\begin{array}{cccccc}
0 & 0  & 0 & 0 & 0 & 1 \\
0 & 0 & 0 & 0 & -1 & 0 \\
0 & 0 & 0 & 1 & 0 & 0\\
0 & 0 & 1 & 0 & 0 & 0 \\
0 & -1 & 0 &  0 & 0 & 0 \\
1 & 0 & 0 & 0 & 0 & 0
\end{array} \right).$$

Below we present the steps of the procedure of computation for the Evans function.
After fixing the values of the parameters corresponding to the
appearance of the homoclinic loop, we choose the starting points
from which we calculate the homoclinic solution. The boundary
conditions at $\pm \infty$ are replaced by the boundary conditions
posed at finite (sufficiently remote) points $\pm T$. The values
$R(\pm T)$ and $Y(\pm T)$ should be chosen with great precision.
They are obtained by solving the linearized system
(\ref{Vladimirov:ds1}). The eigenvectors of the matrix
$M_{lin}=M\left(R_1,\,0  \right)^{tr}$ (see formula
(\ref{Vladimirov:mlins})) are chosen in the form
$$(R(\pm T),Y(\pm T))^{tr}=(R_1,\,0)^{tr}+\varepsilon \vec{q}, $$ where  $\vec
q=\left(1,\,\mp \sqrt{\frac{-2\,E\,R_1-\beta
R_1^3+4s^2}{2R_1^3\gamma}}\right)^{tr}$ are the eigenvectors of the
matrix $M_{lin}$.

\begin{figure}
\begin{center}
\includegraphics[totalheight=1.5in]{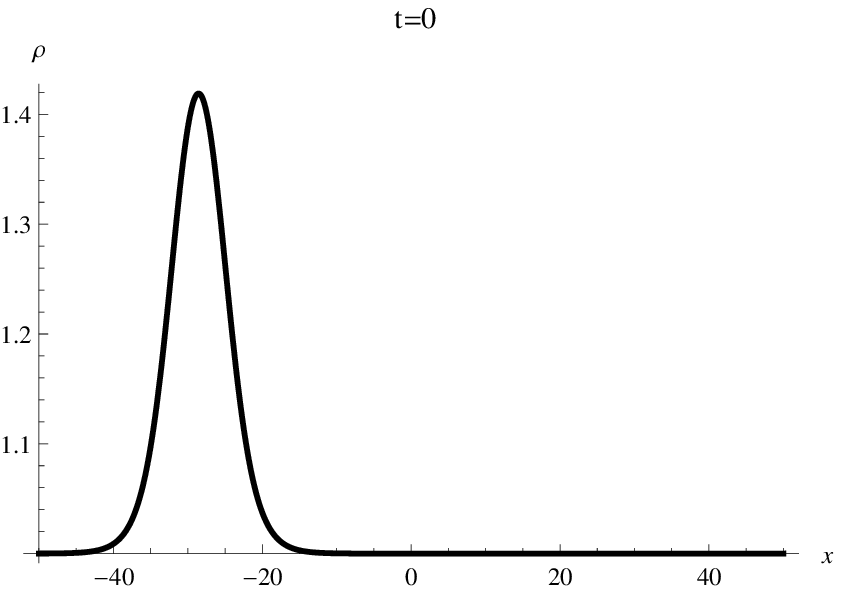}
\includegraphics[totalheight=1.5in]{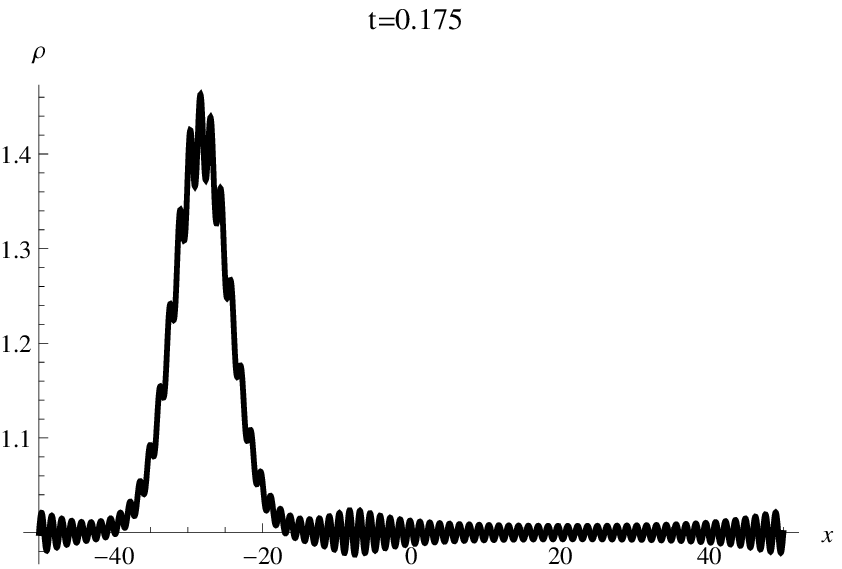}
\\
\includegraphics[totalheight=1.5in]{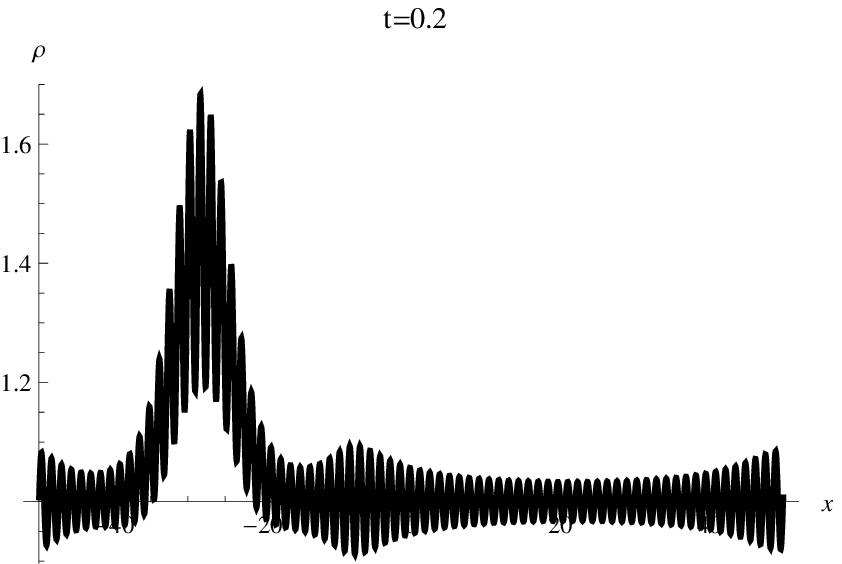}
\includegraphics[totalheight=1.5in]{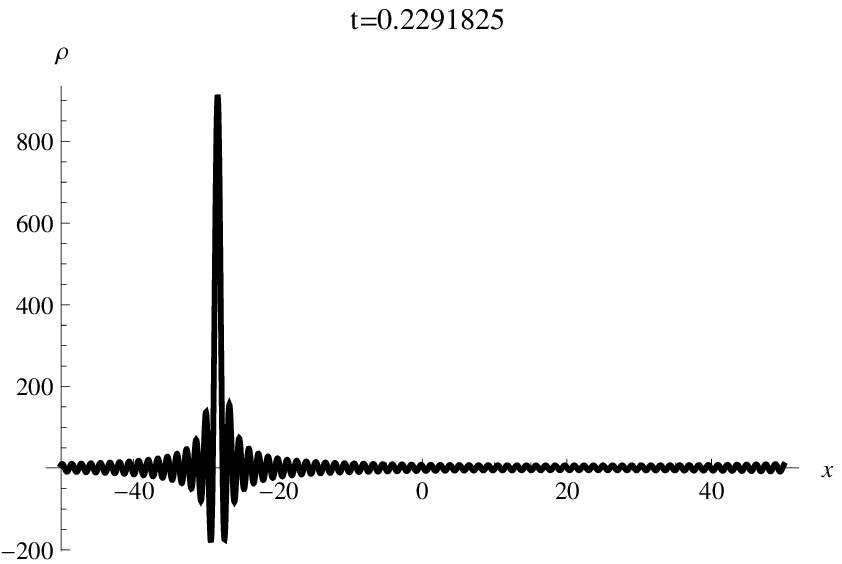}
 \caption{Numerical simulation of system (\ref{Vladimirov:mainspat}) performed with   $R_1=1.0$,
$\nu=0$, $s=1.6$,  $\beta=1.75$  and  $\gamma=3$
}\label{Fig:solitunst}
\end{center}
\end{figure}

The parameter $\varepsilon$  is specified by smoothly sewing the
solutions starting from the initial values $z=-T$ and $z=T$.
For $T=25$ we get  $\varepsilon=1.885\cdot
10^{-6}$.

\begin{figure}
\begin{center}
\includegraphics[totalheight=2.0 in]{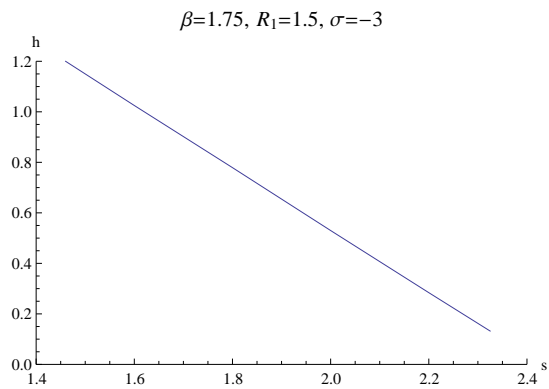}
\caption{The dependence of the maximal depth $h$ of the solitary
wave of rarefaction on the velocity $s$. }\label{Fig:Depth}
\end{center}
\end{figure}

\begin{figure}
\begin{center}
\includegraphics[totalheight=2.0 in]{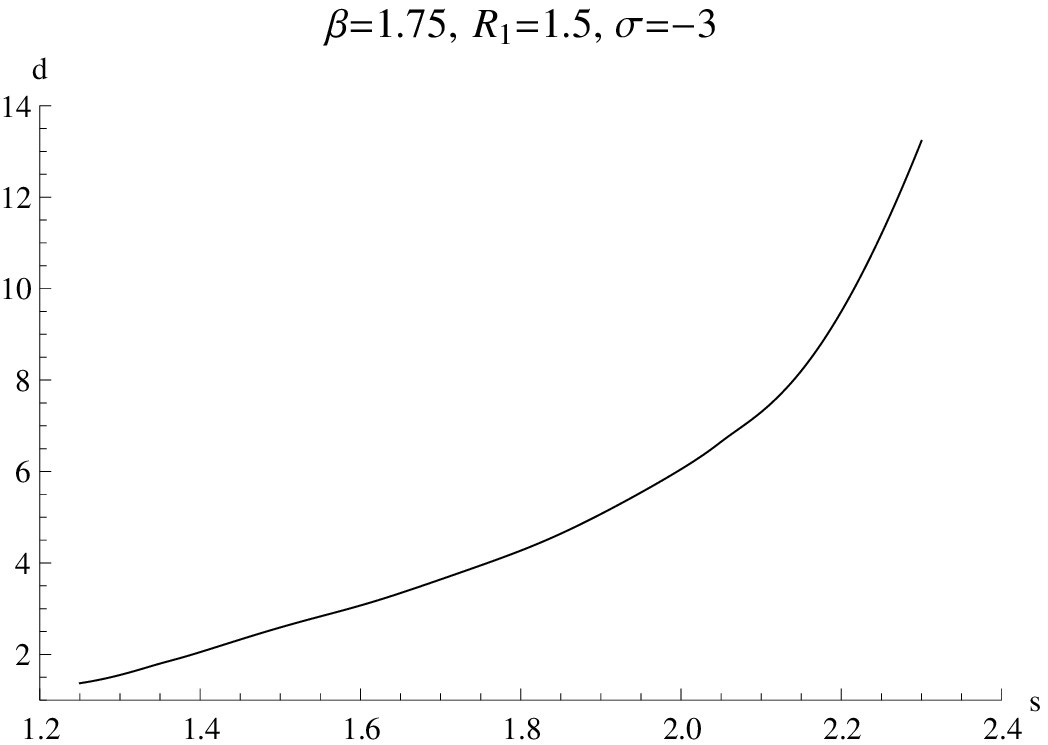}
\caption{The dependence of the effective width $d$ of the solitary
wave of rarefaction on the velocity  $s$.}\label{Fig:Width}
\end{center}
\end{figure}

Next we perform a change of variables \cite{derks}
$$U=\exp\{\mu_{\pm}z\}\tilde U,$$
 that reduces the  system  (\ref{extend})  to the following form:
\begin{equation}\label{lin_cntr}
    \tilde U^\prime=(A^{(2)}-\mu_{\pm} I_4) \tilde U.
\end{equation}
We integrate the system (\ref{lin_cntr}) with appropriate  initial
conditions from  $z=+T$ to $z=+0$, and then from  $z=-T$ to $z=-0$.
The function $E(\lambda)$ is proportional to the $\left\langle
{\tilde U^-}, \Sigma {\tilde U}^+\right\rangle,$ so the zeroes of this
function coincide with the zeroes of the function
$$\tilde E(\lambda)=\left\langle {\tilde U^-}, \Sigma {\tilde
U}^+\right\rangle.$$

The results obtained by the implementation of the above algorithm
are presented below. In order to investigate the behavior of the
Evans function within the domain lying in the positive half-plane,
the Nyquist diagrams $\left(
Re\,E(\lambda),\,\,Im\,E(\lambda)\right)$ are used; they enable
us to fix the number of complex eigenvalues of the linearized
operator $L$. We construct numerically the map $\mathbb{C}\ni
\lambda\,\rightarrow\,E(\lambda)$ for $\lambda$ running along
$\partial\,\Omega$, where $\Omega\in\,C^+$ is a closed set bounded by the straight line $z=a>0$ and the semi-circle $z=b\,\exp{i\,\pi t}$, $-\pi/2<t<\pi/2, \quad 0<a<<1<<b$ \cite{derks,Blank}.

%For $\gamma\,>0$ the solitary wave solutions are unstable  (and this is
%confirmed by the numerical simulation shown at
%Figure~\ref{Fig:solitunst}), so we do not analyze this case.

The
Nyquist diagrams obtained for $\gamma<0$ are shown in
Figures~~\ref{Fig:Nyq1} - \ref{Fig:Nyq2}. The subsequent analysis
shows that the winding numbers in all these cases equal zero, so the
corresponding domains do not contain the values of the spectral
parameter belonging to $\sigma_{discr}$.

\begin{figure}
\begin{center}
\includegraphics[totalheight=1.2 in]{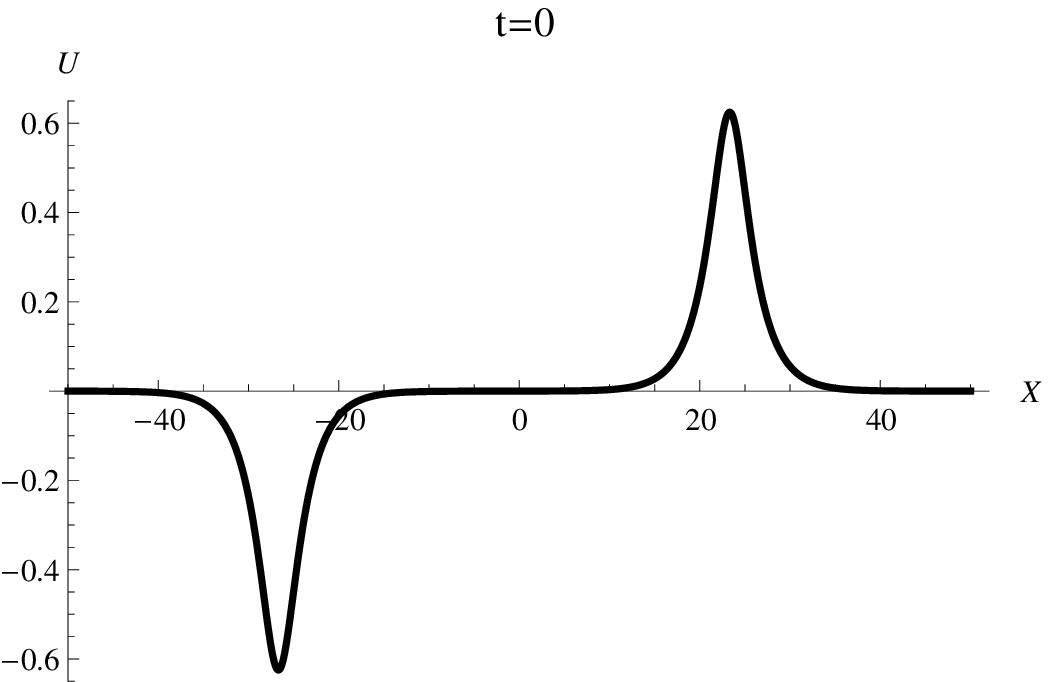}
\includegraphics[totalheight=1.2 in]{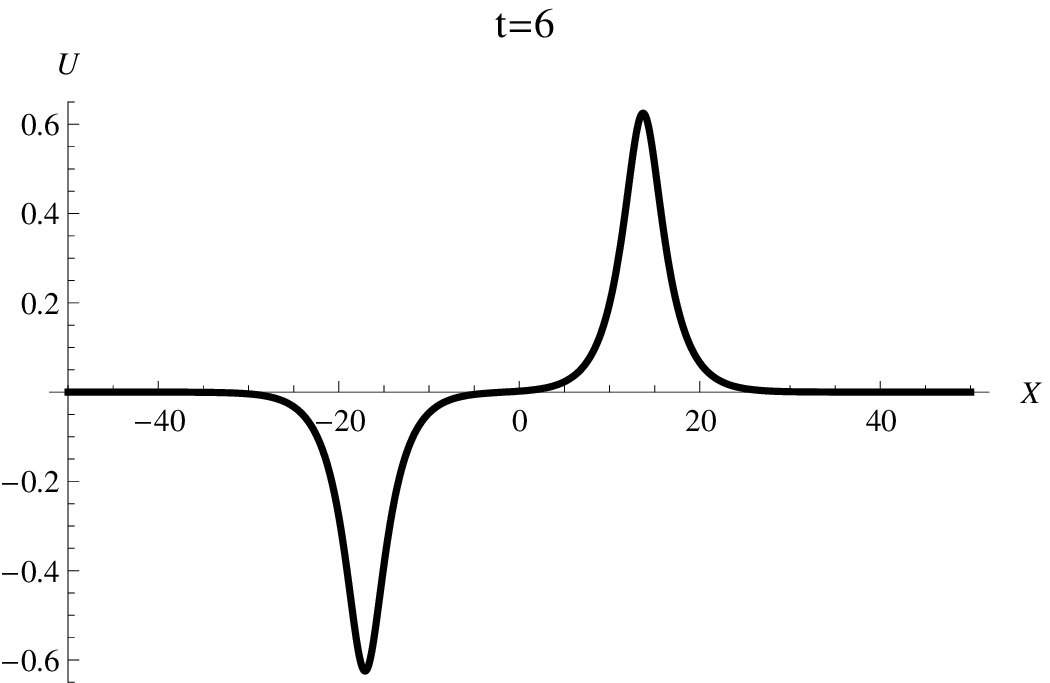}
\includegraphics[totalheight=1.2 in]{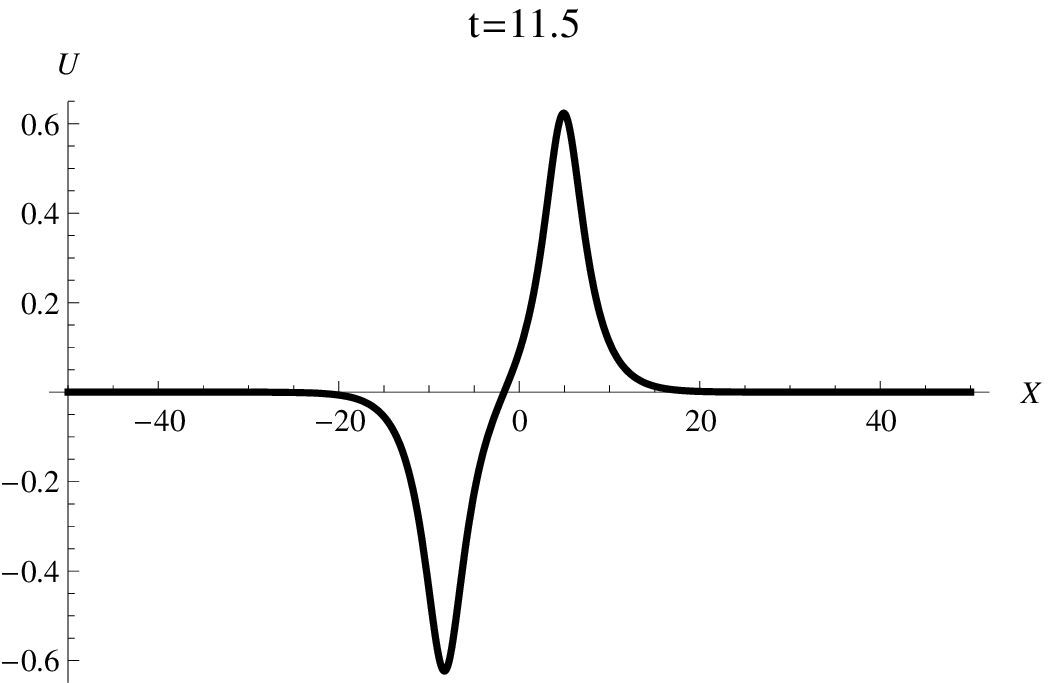}
\\
\includegraphics[totalheight=1.2 in]{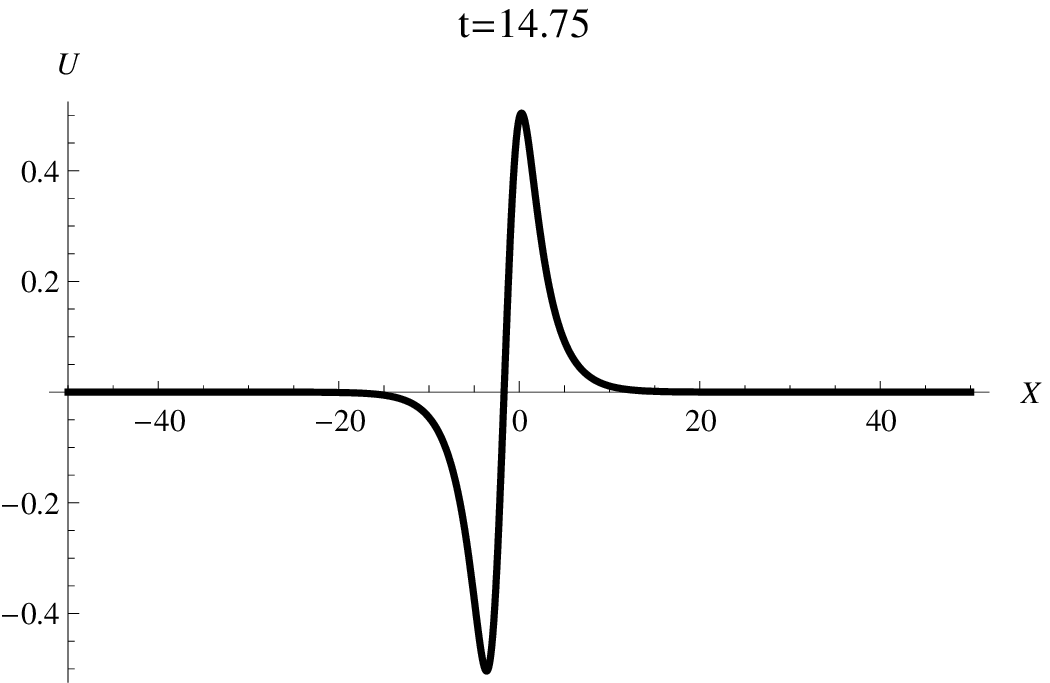}
\includegraphics[totalheight=1.2 in]{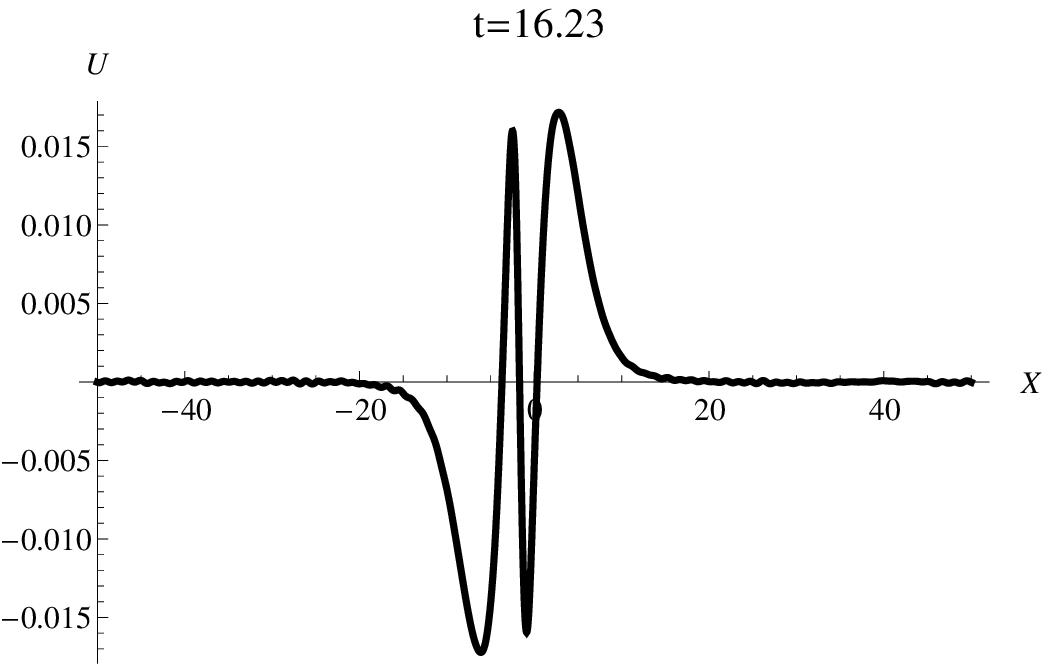}
\includegraphics[totalheight=1.2 in]{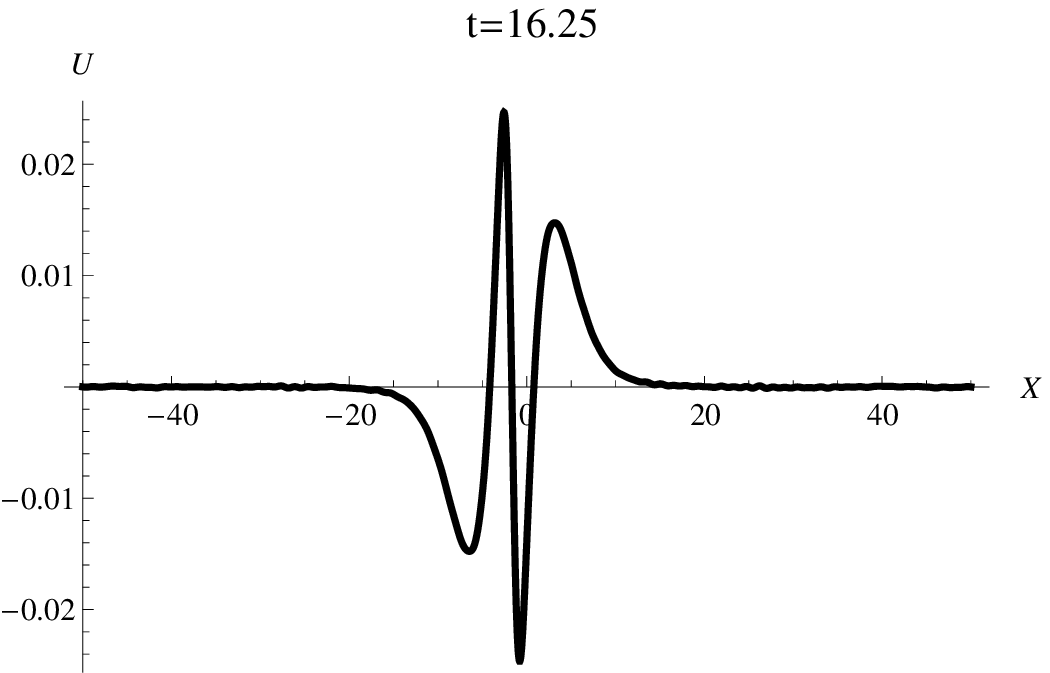}
\\
\includegraphics[totalheight=1.2 in]{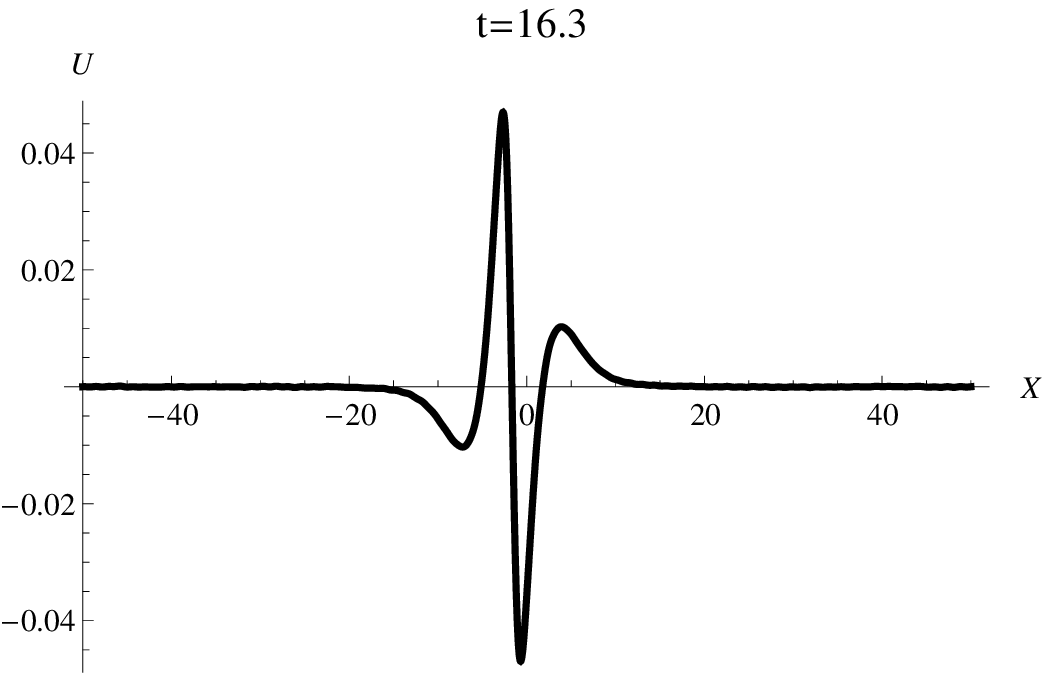}
\includegraphics[totalheight=1.2 in]{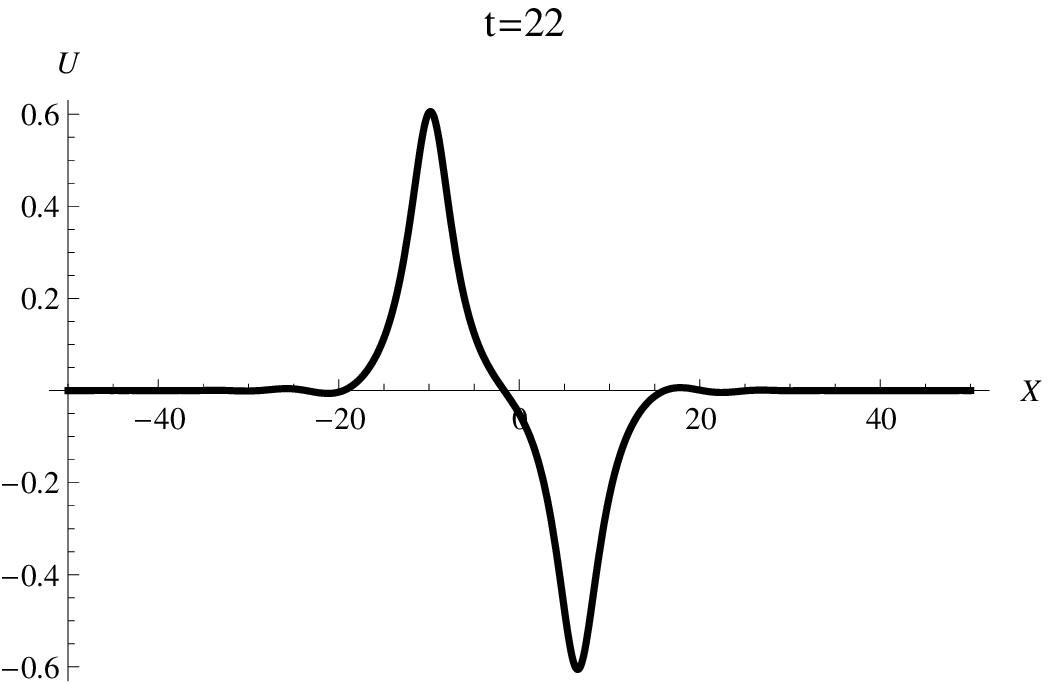}
\includegraphics[totalheight=1.2 in]{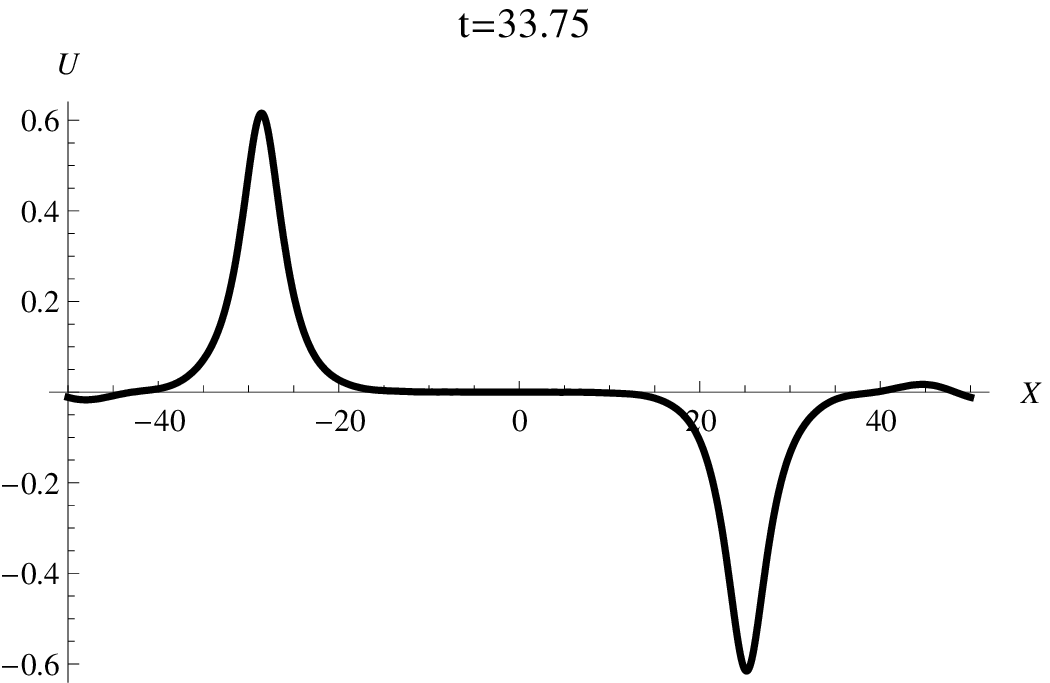}
\caption{Numerical simulation of collision for two solitary
 waves moving towards each other. Calculations are performed for the
  following values of the parameters:    $R_1=1.282,$
$\nu=0$, $s=\,\pm\,1.6$ (for the left and right  perturbation
correspondingly), $\beta=1.75$ and $\gamma=-3$
}\label{Fig:collision}
\end{center}
\end{figure}

\begin{figure}
\begin{center}
\includegraphics[totalheight=2.2 in]{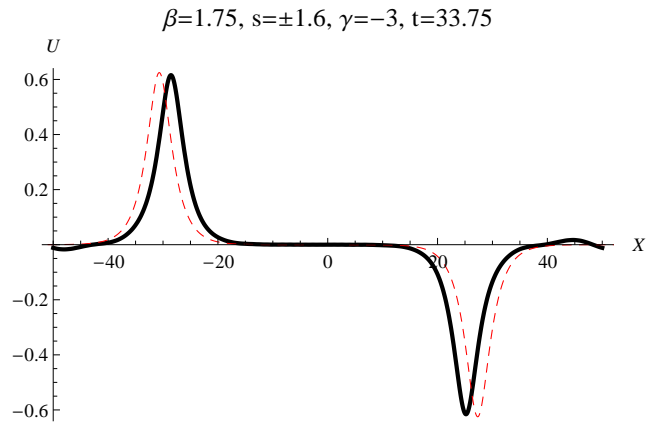}
\caption{A ``negative" phase shift attained by the solitary waves
after the interaction. The dashed line represents the result of the
evolution of unperturbed solitary waves. Calculated for
$R_1=1.282,$ $\nu=0$, $s=\,\pm\,1.6$,
 $\beta=1.75$ and $\gamma=-3$
}\label{Fig:Phashift}
\end{center}
\end{figure}

\begin{figure}
\begin{center}
\includegraphics[totalheight=1.2 in]{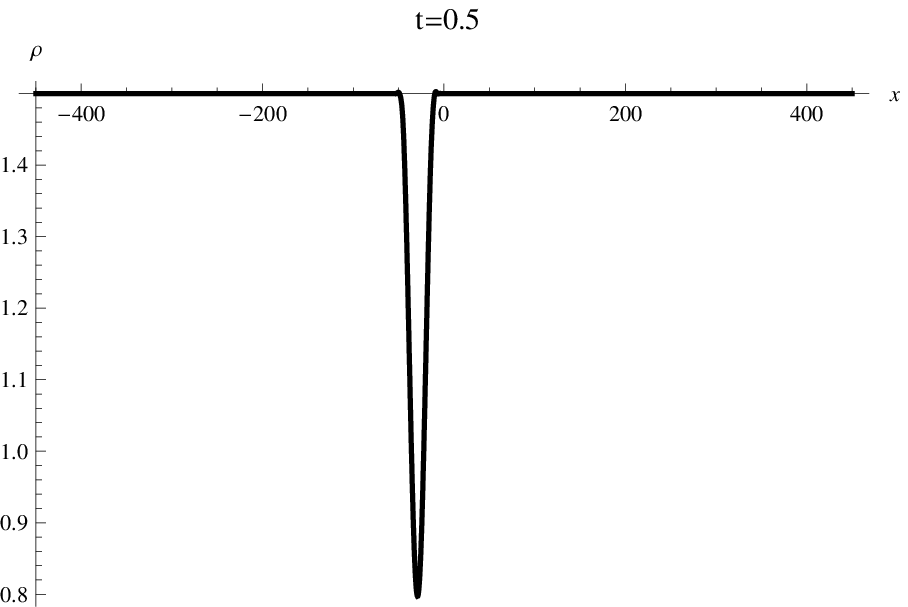}
\includegraphics[totalheight=1.2 in]{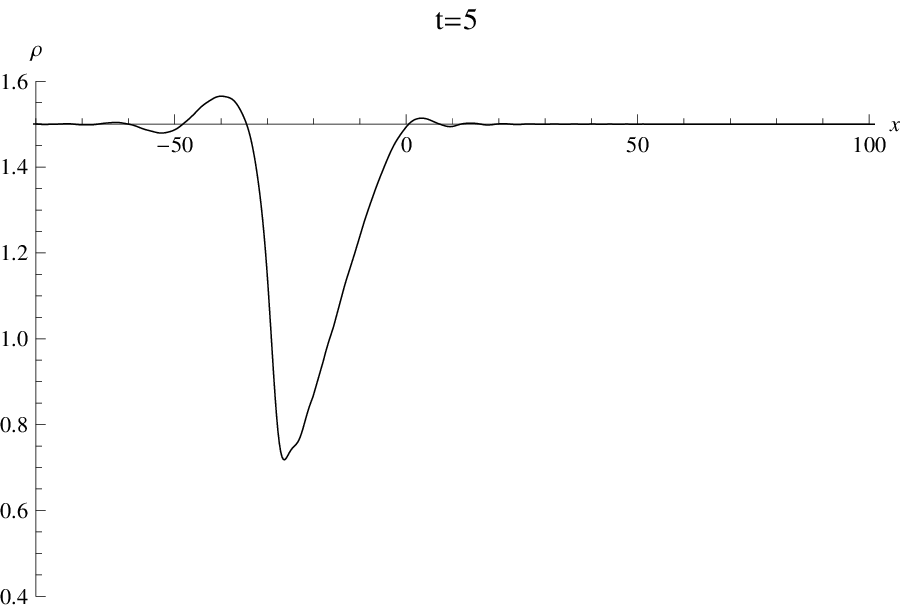}
\includegraphics[totalheight=1.2 in]{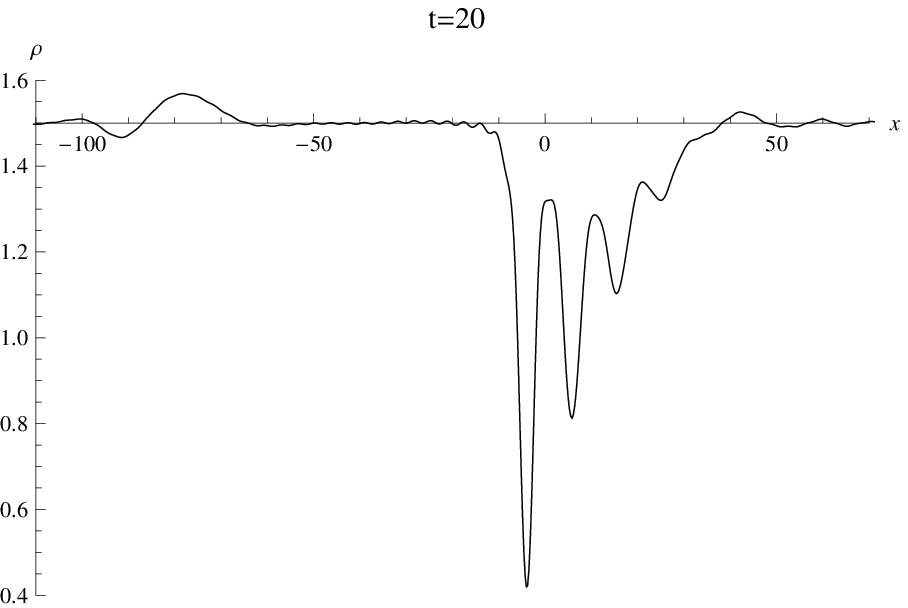}
\vspace{5mm}
\\
\includegraphics[totalheight=1.2 in]{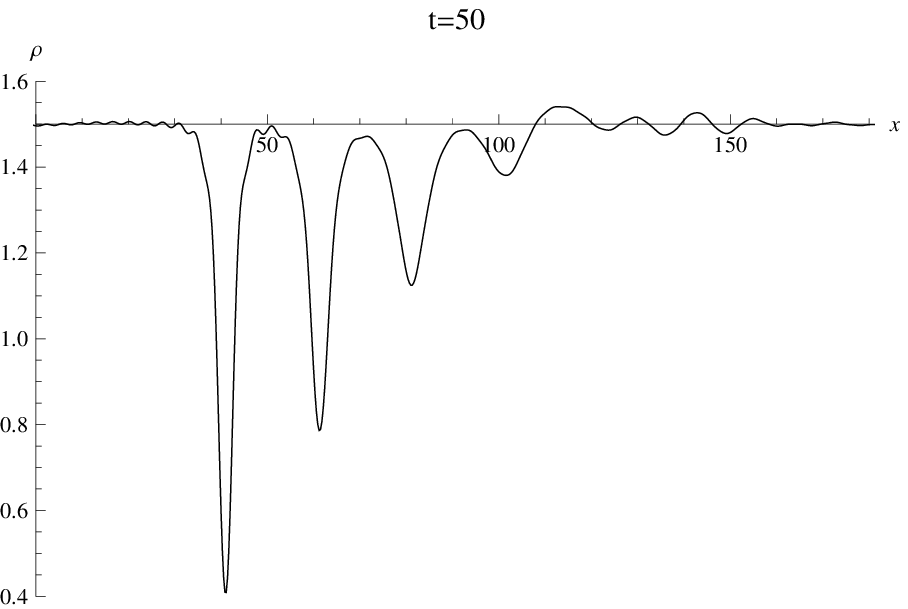}
\includegraphics[totalheight=1.2 in]{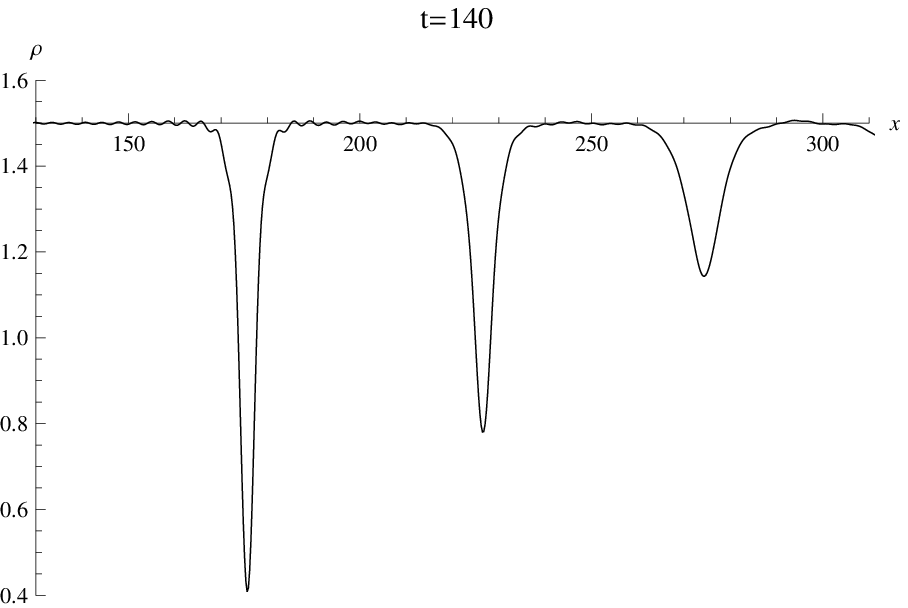}
\caption{Numerical solution of the Cauchy problem for the system
(\ref{Vladimirov:mainspat}) leading to the formation of a train of
solitary waves of rarefaction. Calculations are performed with the
functions (\ref{Multisol}), taken as the Cauchy data, and for the
following values of the parameters:    $R_1=1.5,$ $\nu=0$, $s=1.8$,
$\beta=1.75$, $\gamma=-3$, $A=19$, $B=0.35$ and $x_0=30$.
}\label{Fig:Tsug}
\end{center}
\end{figure}

Now let us describe the results of numerical study of the solitary
waves. The numerical experiments in which the Cauchy problem for
the  system (\ref{Vladimirov:mainspat}) was solved with the
solitary wave solutions taken as the initial data  show that the
solitary wave solutions obtained for $\gamma\,>0$ are unstable (see
Fig.~\ref{Fig:solitunst}), while those corresponding to
$\gamma\,<0$ are stable and  evolve in a self-similar mode. It is
worth noticing that the properties of the solitary waves of rarefaction
supported by the system (\ref{Vladimirov:mainspat}) are different
from those of ``classical" solitons. For example, the maximal depth
$h$ of the solitary wave decreases when the velocity $s$ increases,
see Fig.~\ref{Fig:Depth}. On the contrary, the effective width $d$ of the
wave pack measured at the depth $h/2$ nonlinearly increases as $s$
increases, see Fig.~\ref{Fig:Width}.

\begin{figure}
\begin{center}
\includegraphics[totalheight=1.6 in]{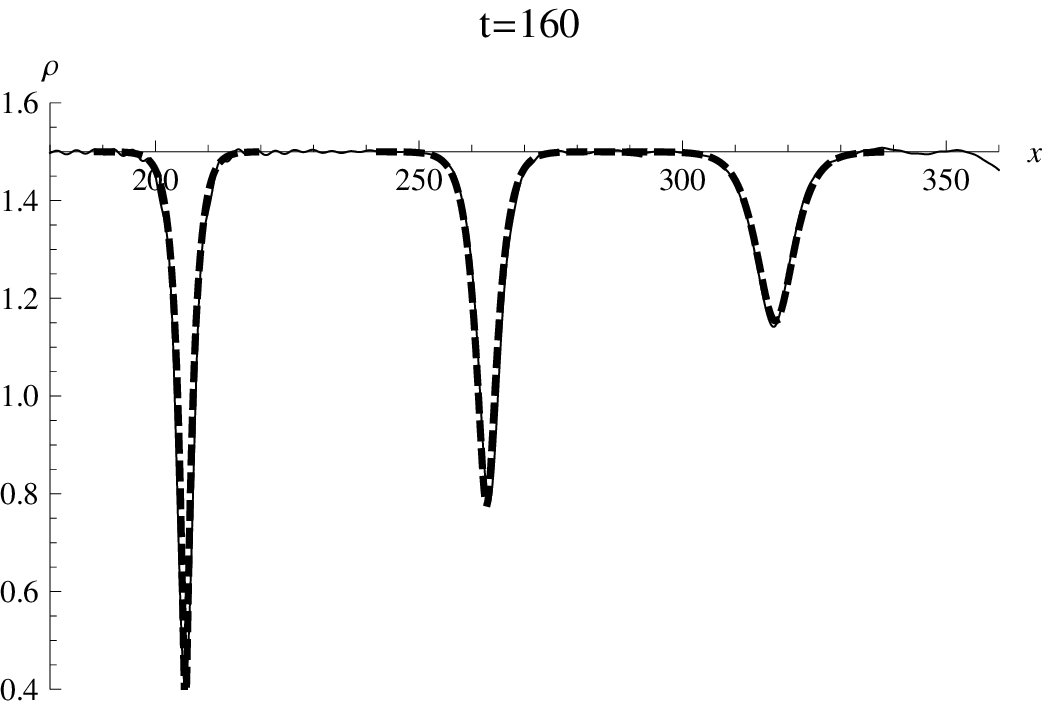}
\caption{Numerical solution of the Cauchy problem for the system
(\ref{Vladimirov:mainspat}) at $t=160$ (solid) on the background of "free" solitary waves (dashed), obtained at the same values of the parameters $R_1,\,\,\gamma,\,\,\nu,\,\,\beta$, but with  $s$ changing from case to case as follows:   $s=1.5$ (left), $s=1.845$ (middle), $s=2.15$ (right)  }\label{Fig:TsugControl}
\end{center}
\end{figure}

 Using the fact that the reduced system does not depend on the sign of
velocity $s$, one is also able to choose as the Cauchy data a pair
of solitary wave solutions separated with suffice spatial interval
and moving toward each other. Numerical simulations show that the
wave packs manifest the soliton behavior, maintaining their shape
after the interaction, see Fig.~\ref{Fig:collision}. Note
that the solitary waves of rarefaction gain the ``negative" phase
shift after the mutual collision, see Fig.~\ref{Fig:Phashift}.

We  are also interested in finding out whether there exists
a set of the initial data
producing a series of solitary waves. In numerical experiments the
initial data of the following form have been used:
\begin{eqnarray}\label{Multisol}
\rho(0,\,x)=\left\{\begin{array}{c} R_1 \quad  \verb"if " \quad
x<-A-x_0, \\
R_1-B\left[\cos(\pi\,\frac{x+x_0}{A}+1) \right] \quad  \verb"if "
\quad -A-x_0 \leq\,x\,\leq\,A-x_0, \\
R_1 \quad  \verb"if " \quad x>A-x_0,
\end{array} \right.   \nonumber \\
u(0,x)=s\left(\frac{1}{R_1}-\frac{1}{\rho(0,\,x)}    \right),
\qquad\qquad\qquad\qquad\qquad\qquad\qquad\qquad\qquad\qquad
\end{eqnarray}
where $s,\,\,A,\,\,x_0$ and $B$ are non-negative parameters.
Numerical experiments reveal the formation of the  series of
solitary waves of rarefaction moving  with different velocities, as
it is illustrated in Fig.~\ref{Fig:Tsug}. The solitons shown at
this figure move from left to right, with the velocities inversely
proportional to their maximal depth, which is in agreement with the
above properties of the individual solitary wave of
rarefaction. It occurs to be possible to choose parameters of ''free '' solitary waves in such a way that, being shifted on a proper distance, they coincide with the solitary waves of rarefaction formed in numerical solutions of the above Cauchy's problem, Fig.~\ref{Fig:TsugControl},

\section{Conclusions and discussion}

In the present paper we have performed an analysis of a hydrodynamic-type system
 (\ref{Vladimirov:mainspat}). In general position the
 system happens to admit only four symmetry generators
and seven local conservation laws. The results of the symmetry
analysis, as well as the study of soliton dynamics suggest that the
complete integrability of the system (\ref{Vladimirov:mainspat}) is
highly unlikely, at least for the generic values of parameters.

Nevertheless, the system is shown to possess a one-parameter family
of stable solitary wave solutions manifesting some features of
"true" solitons. Let us stress that existence of a one-parameter
family of solitary wave solutions is connected with the employment of the
dynamic equation of state (\ref{Eq:tspat}), taking account of the effects
of pure spatial non-locality. The presence of one-parametric families
of localized TW solutions is rather not inherent to another known hydrodynamic-type
non-local models \cite{Danbook}, accounting for temporal \cite{Vlad2008,Cybcon,VlaMah} or spatio-temporal \cite{VlaSk2000} non-localities.
Note that throughout the text we do not use the
analytical description of the solitary wave solutions. Their
existence is proved on the basis of the qualitative study of the
associated dynamical system. Such study is more relevant and
informative than the attempts to find an analytic description for
solitary waves, for they are rarely expressed in terms of elementary
(or even special) functions. Moreover,
 applying the qualitative methods enables one to cover the whole set
 of invariant solutions belonging to the given family.

The main goal of this work is the study of spectral stability for
solitary waves and the dynamical features thereof. A stable evolution of
the solitary waves of compression corresponding to the case of
$\gamma>0$ is virtually impossible because the intersection of
$\sigma$ with $\mathbb{C}^+$ is nonempty.

In the case of $\gamma<0$ the spectrum of linearized problem happens
to  lie in the set $\mathbb{C}\backslash\,\mathbb{C^+}$. This
conclusion is made on the basis of the analytic investigation of
the continuous spectrum and numerical study of the Evans function
performed for the selected values of parameters and covering a
sufficiently large domain of the half-plane $\mathbb{C}^+$.  These results
are further backed by the estimate of the Evans function made for
large $|\lambda|$ lying in the positive half-space.

 %Let us stress in conclusion that the number of the
 %conserved quantities admitted by the system
 %(\ref{Vladimirov:mainspat}) does not depend on the sign of $\gamma$
 %so this number seems to be of no crucial significance for the stability properties. However
 %it is worth noting that the functional behavior of  the conserved  densities $R_3,\,R_4,\,R_6$ and $R_7$
 %is strongly dependent on the sign of the parameter $\gamma$.

 A remarkable and somewhat unexpected result obtained in the numerical experiments is
 the elastic dynamics of interaction of solitary wave solutions. Let
 us stress that a mere possibility to perform such experiments is rare,
 for the soliton-like TW solutions supported by non-integrable evolutionary equations occur, as a
 rule, for specific values of parameters, including the magnitude and direction of the velocity.

 The numerical studies reveal some peculiarities of the solitary waves of
 rarefaction, in particular, an anomalous dependence of the depth and
 the effective width upon the wave pack velocity. These features are
 well understood on the basis of qualitative analysis. The decrease
 of the depth of the solitary wave with the increase of the velocity is
 related to the properties of solutions to the algebraic equation
 (\ref{Vladimirov:cp_spat}) determining the
 horizontal coordinate of the stationary point $(R_2,\,0)$. Thus,
 when the velocity $s$ obeying the inequalities (\ref{gammin}) increases, the point
 $R_2$ in which the straight line $E\,R-s^2$ intersects the  curve
 $\frac{\beta}{\nu+2}\,R^{\nu+3}$ moves towards $R_1$. Hence, the
 maximal depth of the solitary wave, which is proportional to
 $|R_2-R_1|$, decreases. The simultaneous increase of the effective width of the
 solitary wave is related to the fact that all the points of the corresponding homoclinic
 trajectory lie in a vicinity of the stationary points, hence the
 phase velocity is small in every point of the homoclinic loop and
 tends to zero as $s$ approaches $\sqrt{\beta\,R_1^{\nu+3}}$ from below.

 When
 $s$ approaches $\sqrt{\frac{\beta(\nu+1)}{\nu+2}\,R_1^{\nu+3}}$ from above,
 a different effect is observed. Since the homoclinic loop is locked
 from the left by the vertical axis, it attains the shape of an
 equilateral triangle, symmetric w.r.t. the horizontal axis (see  Fig.~~\ref{Fig:ris1}~b).
 Its  base approaches the vertical axis, as the parameter
 $s$ decreases, thus resulting in an unlimited growth of the phase
 velocity in the region corresponding to the bottom of the solitary
 wave of rarefaction. This, in turn, causes the creation of an abruptly narrowing
 solitary wave. In fact, the solitary wave solution turns into a spike
 as $s$ approaches the left critical value, which is clearly seen
 in numerical experiments.

%A   study of the interaction of wave packs  becomes
%possible in our case thanks to the  stability of a single solitary wave solution.
% Another related factor  is that the reduced system is invariant
%under the sign of the velocity of solitary wave, which thus can move
%in both directions.

As a final remark, note that, in addition to obtaining the
rigorous results concerning the spectral stability of the TW
solution, the list of open problems for  the system
(\ref{Vladimirov:mainspat}) includes the issue of more general
stability and attractive features
\cite{Barenblatt,Kamin_Rosenau1,Kamin_Rosenau2}, the study of
interaction of the waves for the wide range of values of parameters,
and special treatment of the distinguished special case $\nu=1$, for which the
system (\ref{Vladimirov:mainspat}) possesses a nonlocal symmetry.

\section*{ Appendix 1}

 \vspace{5mm}

Following a common practice \cite{shoot}, we shall calculate the
Evans function for a system of ODEs with constant coefficients
approximating the linearized system (\ref{dyn_syst}). Let us employ
a trigonometric representation
$\lambda=|\lambda|\,e^{i\,\varphi}$ for $\lambda$, assuming that $|\lambda|>>1$,
and $-\frac{\pi}{2}<\varphi<\frac{\pi}{2}$. Applying the scaling
transformation
\[
\bar g=\frac{g}{R}, \qquad  \bar \eta = |\lambda|^{-1/2}\,\eta,
\qquad \bar \chi = |\lambda|^{-1}\,\chi, \qquad \bar
f=|\lambda|^{-1/2}\,f,
\]
and passing to the new independent variable
$\frac{d}{d\,\tau}=\frac{R}{|\lambda|^{1/2}}\,\frac{d}{d\,z}$, we
obtain for $|\lambda|>>1$ an approximate system
\begin{equation}\label{ApproxFODE}
\frac{d}{d\,\tau}\left(  \begin{array}{c}\bar g \\\bar\eta \\
\bar \chi
\\ \bar f
\end{array} \right)=\tilde M\,\left(  \begin{array}{c}\bar g \\\bar\eta \\
\bar \chi
\\ \bar f
\end{array} \right)= \,\left(\begin{array}{cccc} 0 & 1 & 0 & 0 \\
0 & 0 & R_0 & 0 \\ 0 &  0  & 0 & \frac{e^{i\,\varphi}}{\gamma} \\
e^{i\,\varphi} & D_0& 0 & 0
\end{array} \right) \left(  \begin{array}{c}\bar g \\\bar\eta \\
\bar \chi
\\ \bar f
\end{array} \right),\end{equation}
where $R_0=R(0).$

The matrix $\tilde M$ has four distinct eigenvalues
\[
\vec V_k=\left(
R_0,\,\rho\,R_0\,e^{i\,\alpha_k},\,\rho^3\,e^{2\,i\,\alpha_k},\,\gamma\,R_0^4
\,e^{i\,(3\alpha_k-\varphi)} \right)^{tr},\] where $\rho=\left(
R_0/|\gamma|\right)^{1/4}$,
\[
\alpha_k=\frac{\varphi+(k-1)\,\pi}{2}, \qquad k=1,...4.
\]
Thus, in the generic case two eigenvectors belong to the unstable
invariant manifold, and two other to the stable one. So the Evans
function for the asymptotic problem is proportional to
\begin{equation}
D= R_0^5\, \det \left[\begin{array}{cccc} 1 & 1 & 1 & 1 \\
e^{i\,\alpha_1} & e^{i\,\alpha_2} & e^{i\,\alpha_3} &
e^{i\,\alpha_4} \\ e^{2\,i\,\alpha_1} & e^{2\,i\,\alpha_2} &
e^{2\,i\,\alpha_3} & e^{2\,i\,\alpha_4} \\
e^{i(3 \alpha_1-\varphi)} & e^{i(3 \alpha_2-\varphi)} & e^{i(3
\alpha_3-\varphi)} & e^{i(3 \alpha_4-\varphi)}
 \end{array}  \right].\end{equation}
Upon performing straightforward but tedious
calculations we finally obtain that
\[
D=2\,i\,\,R_0^5\, e^{3\,i\,\varphi/2}
\left(e^{i\,\alpha_2}-e^{i\,\alpha_1} \right)
\,\left(e^{i\,\alpha_3}-e^{i\,\alpha_1}
\right)\,\left(e^{i\,\alpha_4}-e^{i\,\alpha_1} \right).
\]
Further analysis shows that under the above restrictions this number
never vanishes, which indicates that the Evans function is nonzero
for  large $\lambda$ with positive real part.

\section*{ Appendix 2}

\vspace{2mm}

Consider the eigenvalues of the matrix $A_\infty$ corresponding to
the small values of $\lambda$. The characteristic equation    $\det
\left[A_\infty-\mu \,I\right]=0$  can be written as follows:
\begin{equation}\label{char}
    \mu^4+\alpha_2 \mu^2+ \alpha_1 \mu+\alpha_0=0,
\end{equation}
where $\displaystyle \alpha_2=\frac{\beta R_1^3 - s^2
}{R_1^3\gamma}$, $\displaystyle \alpha_1=\frac{2 s \lambda}{R_1^3
\gamma} $, $\displaystyle \alpha_0=-\frac{\lambda^2}{R_1^3 \gamma}$.
In order to analyze the behavior of the eigenvalues for small
$\lambda\not = 0$, we represent the solutions of (\ref{char}) in the
form of series
$$
\mu=\mu_0+\lambda \mu_1+\dots
$$
In the zero-order approximation we obtain the equation
$$\mu_0^2\left(\mu_0^2+\alpha_2\right)=0.$$
This equation has a solution $\mu_0^{1,2}=0$ of multiplicity 2,
and a pair of nonzero solutions $\mu_0^{3,4}=\pm\sqrt{-\alpha_2}$.
For $\alpha_2<0$ or $\beta R_1^3 - s^2>0$ the quantities $\mu_0^{3,4}$ are
real and have different signs. For $\alpha_2>0$ or $\beta R_1^3 -
s^2<0$, they are purely imaginary.

The asymptotic series corresponding to nonzero $\mu_0^{3,4}$ has
the form:
$$
\mu=\mu_0+\frac{s}{\beta R_1^3-s^2}\lambda - \frac{2s^2+\beta
R_1^3}{2(s^2-\beta R_1^3)^2 \mu_0} \lambda^2+\dots
$$
For  $\mu_0^{3,4}$ the expansion is different:
$$
\mu=\lambda \mu_1+\lambda^3 \mu_3+\dots,$$ where $\displaystyle
\mu_1=\frac{1}{s\pm\sqrt{\beta R_1^3}}$, $\displaystyle
\mu_3=\frac{\mu_1^4 R_1^3 \gamma}{-2s+2s^2\mu_1-2\beta\mu_1R_1^3}$.

When  $\alpha_2<0 \rightarrow s>\sqrt{\beta R_1^3}$  and
$\lambda>0$, then the real roots are shifted  to the right, while
the zero roots give rise to a pair of real roots having different signs.
Hence, the system (\ref{dyn_syst}) has a two-dimensional unstable
invariant manifold. When $\alpha_2>0  \rightarrow s<\sqrt{\beta
R_1^3},$ a pair of positive roots is created from zero ones, while
the pair of a pure imaginary roots gain negative real part. Thus, in
this case the system  (\ref{dyn_syst}) has two-dimensional unstable
invariant manifold as well.

\section*{Acknowledgements}

{The authors gratefully acknowledge support from the Polish Ministry
of Science and Higher Education (VV, CM), and from the Ministry of
Education, Youth and Sport of the Czech Republic under RVO funding
for I\v{C}47813059 and from the Grant Agency of the Czech Republic
(GA \v{C}R) under grant P201/11/0356 (AS).\looseness=-1 }

%% The Appendices part is started with the command \appendix;
%% appendix sections are then done as normal sections
%% \appendix

%% \section{}
%% \label{}

%% References
%%
%% Following citation commands can be used in the body text:
%% Usage of \cite is as follows:
%%   \cite{key}          ==>>  [#]
%%   \cite[chap. 2]{key} ==>>  [#, chap. 2]
%%   \citet{key}         ==>>  Author [#]

\end{document}